\documentclass{ws-ijmpe}
\usepackage[super,compress]{cite}
\usepackage[usenames,dvipsnames]{color}
\usepackage{nicefrac}
\usepackage{float,ulem}
\usepackage{epsfig,epstopdf,amsmath,amssymb}

\usepackage{xcolor}
\usepackage{subfig}
\newcommand{\be}{\begin{equation}}
	\newcommand{\ee}{\end{equation}}
\newcommand{\ba}{\begin{eqnarray}}
	\newcommand{\ea}{\end{eqnarray}}

\newcommand{\bas}{\begin{eqnarray*}}
	\newcommand{\eas}{\end{eqnarray*}}

\usepackage{scalerel}

\usepackage{soul}
\usepackage{hyperref}
\usepackage{wrapfig}

\usepackage{slashed}
\usepackage{comment}
\usepackage{float}
\usepackage{graphicx}
\begin{document}
\setcounter{tocdepth}{1} 

\markboth{A. Sarkar et al.}{Dynamics of Hot QCD Matter 2024 - New facilities and instrumentation}

\catchline{}{}{}{}{}

 \title{Dynamics of Hot QCD Matter 2024 - New facilities and instrumentation}

\author{
Amal Sarkar$^{1}$\footnote{amal@iitmandi.ac.in}~,
Prabhakar Palni$^{1}$\footnote{prabhakar@iitmandi.ac.in}~,
Santosh K. Das$^{2}$\footnote{santosh@iitgoa.ac.in}~,
Jaideep Kalani$^{1}$,
Ganesh J. Tambve$^{3}$,
Saptarshi Datta$^{3}$,
Saloni Atreya$^{4}$,
Sachin Rana$^{1}$,
Md Kaosor Ali Mondal$^{1}$,
Poojan Angiras$^{1}$,
Anusree Vijay$^{4}$,
Ganapati Dash$^{4}$,
Prafulla Kumar Behera$^{4}$,
Deepak Samuel$^{5}$,
Theertha C$^{4}$,
(authors)\footnote{Contributors are responsible only for the sections they cosign. Collaboration occurred exclusively among those listed together in the header of each section.}}
\address{
$^{1}$ School of Physical Sciences, Indian Institute of Technology Mandi, Himachal Pradesh, India\\
$^{2}$ School of Physical Sciences, Indian Institute of Technology Goa, Ponda-403401, Goa, India\\
$^{3}$Center for Medical and Radiation Physics, NISER, Bhubaneswar, Jatni - 752050, India\\
$^{4}$ Indian Institute of Technology Madras, Chennai, Tamilnadu 600036, India\\
$^{5}$ Central University of Karnataka, Kalaburagi - 585367, Karnataka, India
}%
\maketitle
\begin{abstract}
This part of the conference proceeding provides a detailed overview of cutting-edge advancements in detector technologies, focusing on their optimization, characterization, and applications in particle physics experiments. Building on the insights and developments presented at the Hot QCD Matter 2022 conference \cite{hotQCD22}, this section of the Hot QCD Matter 2024 proceedings highlights significant advancements in detector technologies. The development of Low Gain Avalanche Diodes (LGADs) into Ultra-Fast Silicon Detectors is explored, demonstrating their potential for superior timing resolution in future high-energy experiments. Simulation studies of Micropattern Gaseous Detectors (MPGDs), including MICROMEGAS and Gas Electron Multiplier (GEM) detectors, provide insights into their performance under high-radiation environments using tools like ANSYS and GARFIELD$^{++}$. A novel GEM foil geometry is proposed for improved gain and durability. Characterization of semiconductor detectors, such as Monolithic MALTA pixel detectors and CMS prototype silicon sensors, is also presented, highlighting their radiation tolerance, imaging capabilities, and structural integrity. These studies underscore the critical role of silicon sensors in ensuring detector reliability and performance. Additionally, the J-PARC muon g-2/EDM experiment is reviewed, showcasing its precision measurements to test Standard Model predictions and explore potential physics beyond. 
By addressing the interplay between detector development, simulation, and characterization, this proceeding showcases a collective effort toward advancing detector technologies and their pivotal role in pushing the boundaries of modern particle physics.
\end{abstract}

\ccode{PACS numbers:12.38.-t, 12.38.Aw}

\tableofcontents 
\section{Optimisation of Low Gain Avalanche Diodes into Ultra Fast Silicon Detectors for Future Particle Physics Experiments}
\author{Jaideep Kalani, Saptarshi Datta, Ganesh J. Tambave, Prabhakar Palni}

\bigskip

\begin{abstract}
Future collider experiments require tracking detectors with much better (than the existing, about 40 ps) time and spatial resolution. Low-gain avalanche diodes (LGADs) provide a viable solution to achieve a time resolution better than 20 ps along with a spatial resolution of $< 50$ µm using pixel segmentation. At HL-LHC, we expect an average of 1.6 collisions/mm.  In this work, we studied the correlation of different configurations of LGAD and irradiation parameters with the signal output from electronic read-out associated with the LGADs. The WeightField2 simulation package is used to characterize detector designs. We studied the variation of Bias Voltage, Gain implant layer concentration, and sensor thickness to obtain the criterion for optimal signal output, particularly for n-in-p monolithic LGAD of p-doped Si bulk. Taking into account practical usage, we determined the desired sensor bulk width. In this work, we focus on ultra thin sensors of thickness less than 50 µm to lower the time till the signal decay. The investigation showed 20 µm thick sensors to be efficient in obtaining a considerable gain. Simulations are run for room temperature to account for radiation damage, gain quenching, and other lattice defects associated with semiconductor bulks. We identified LGAD designs with a time resolution ($\sigma_t$) of the order of 20 ps, and this work aimed to characterize the LGAD model over varying bias voltage, thickness, and gain implant  layer concentration.

\end{abstract}





\subsection{Introduction}
LGADs or Low Gain Avalanche Diodes are generally silicon-based solid-state detectors of PIN diode, with an added gain implant layer. It has a moderate gain between 5 to 50. Due to their structure, they have extra-ordinary timing capabilities (time resolution can reach down to 20 ps). LGADs are aimed for use in Time of Flight particle identification (TOF PID ) due to their high charge collection, and ensures timing resolution which is far superior to other detector types. Moreover, they can also withstand high fluence being in the inner barrel and far forward region of the detector line-up and still maintain undiminished timing capabilities. This makes it a promising candidate for future High Luminosity LHC projects. They were first developed by CNM (Centro Nacional de Microelectrónica) as part of the RD50 common project. The most common kind of LGAD used is n-in-p type, figure 1,  which has an n++ layer below the electrode in the p-type bulk. The gain implant is p+ type which creates a strong electric field, $\sim$300kV/cm, at the n++/p+ junction.\cite{cartiglia2015}

The collider experiments, such as at the LHC, are going to work in very high luminosity (HL-LHC), which can be up to 5 to 7 times the nominal luminosity. This huge rise in the luminosity will cause the pile-up to increase up to 150-200 per bunch crossing. Also in such great luminosity, the exposure to radiations (mainly hadrons) will be tremendous to the detectors, which will eventually degrade it. So, LGAD based detectors are planned to to be used for their timing capabilities, and radiation hardness.\cite{rd50} 
\begin{figure*}[ht]
    \centering
    \includegraphics[width=9.0cm]{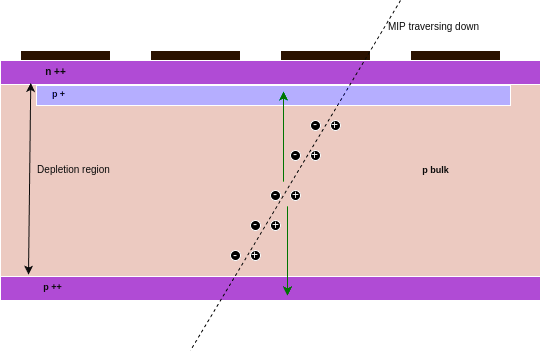} 
    \caption{Schematic diagram of an LGAD with the gain implant layer}
    \label{LGAD_fig}
\end{figure*}
\subsection{Results and Discussions}
All the plots in this study is done using the software WeightField2. This software simulates the LGAD and has features to study current, electric fields, doping, thickness, implant type, radiation hardness and more.

All the plots are drawn based on the simulation of an MIP traversing through an n-in-p type Si-bulk LGAD. Various kinds of studies based on temperature, thickness, fluence, etc. are being done.
\begin{figure*}[ht]
    \centering
    \begin{minipage}{0.45\textwidth}
        \centering
        \includegraphics[width=1.18\linewidth]{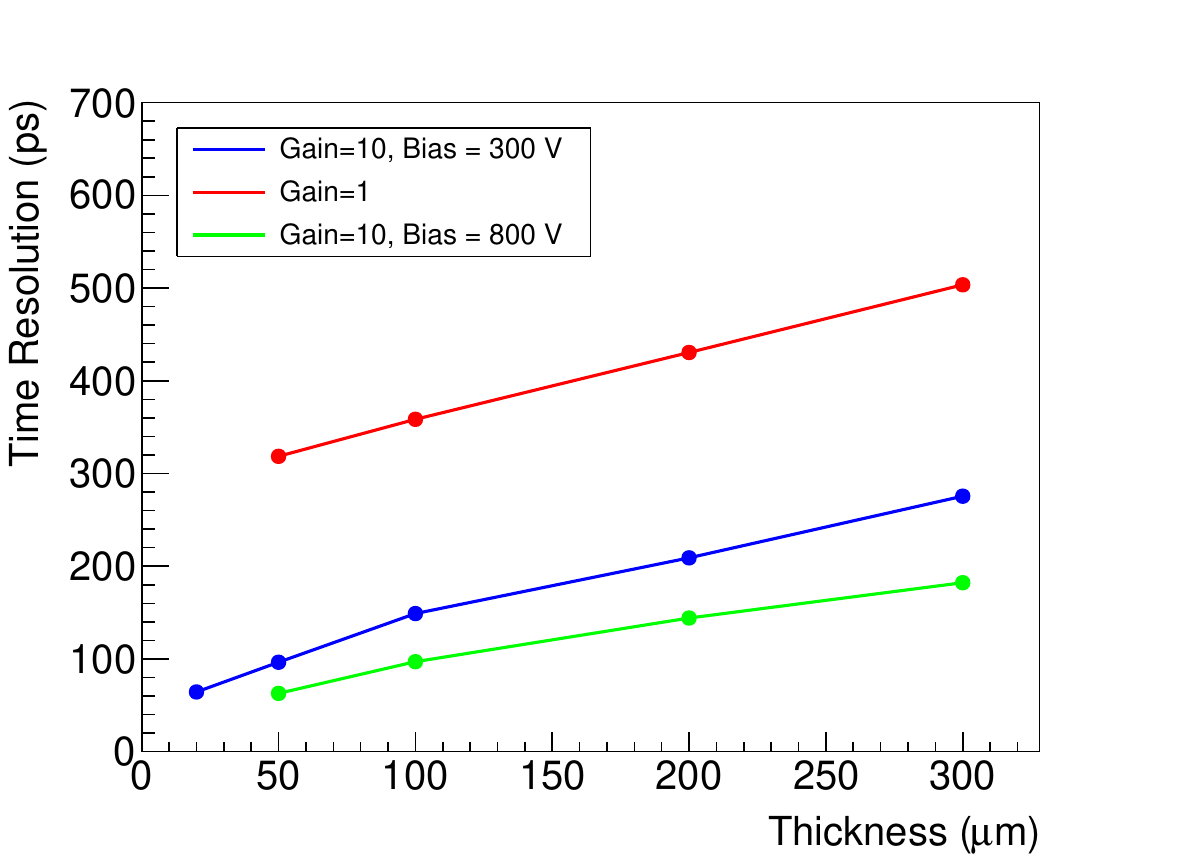}
        \caption{Variation of time resolution of PIN diode and LGAD with sensor bulk thickness}
        \label{time_thick}
    \end{minipage}\hfill
    \begin{minipage}{0.45\textwidth}
        \centering
        \includegraphics[width=1.18\linewidth]{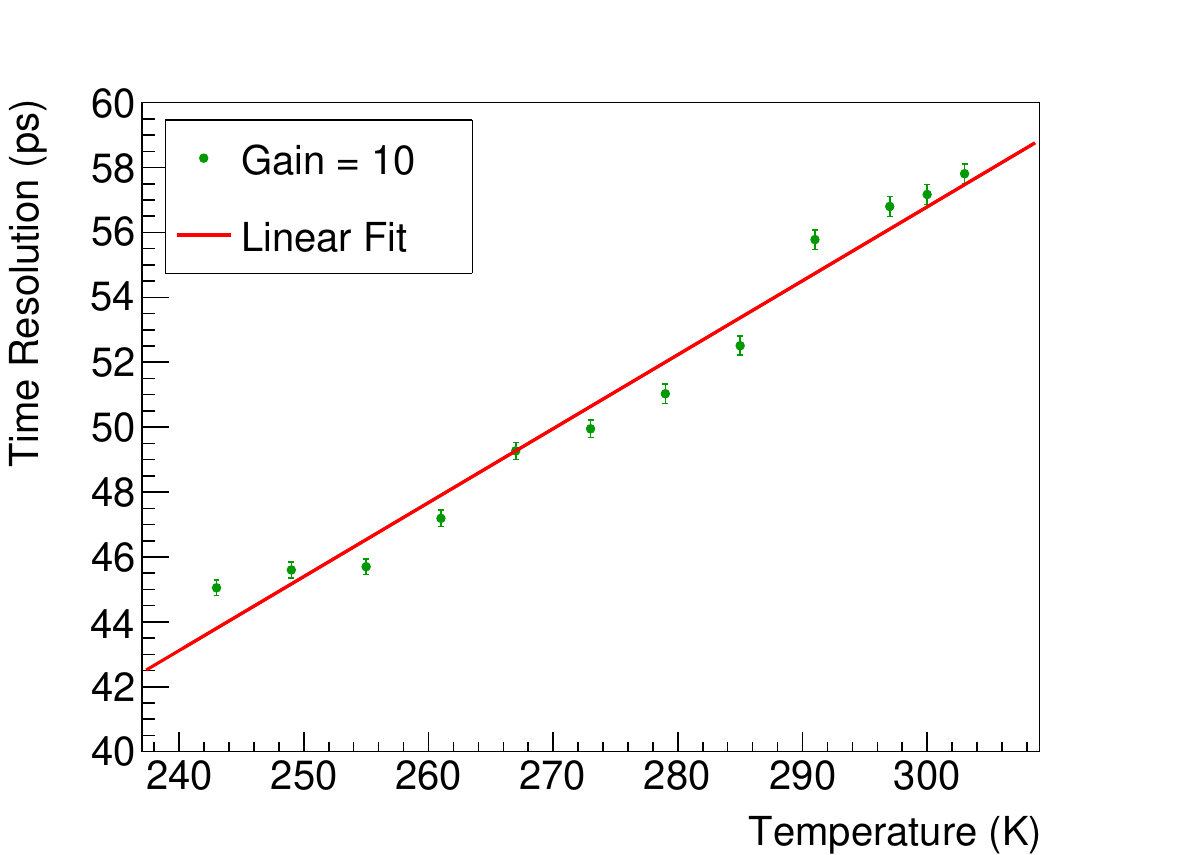}
        \caption{Time resolution as a function of temperature of the sensor at no fluence}
        \label{time_temp}
    \end{minipage}
\end{figure*}

\subsubsection{Thickness Studies}
The figure 2 shows the variation of time resolution with detector bulk thickness for PIN diode and LGAD with gain value 10 at bias voltage 300 V and 800 V. The simulation is throughout done at 300 K for a temperature-independent study. 
\begin{figure*}[ht]
    \centering
    \begin{minipage}{0.45\textwidth}
        \centering
        \includegraphics[width=1.18\linewidth]{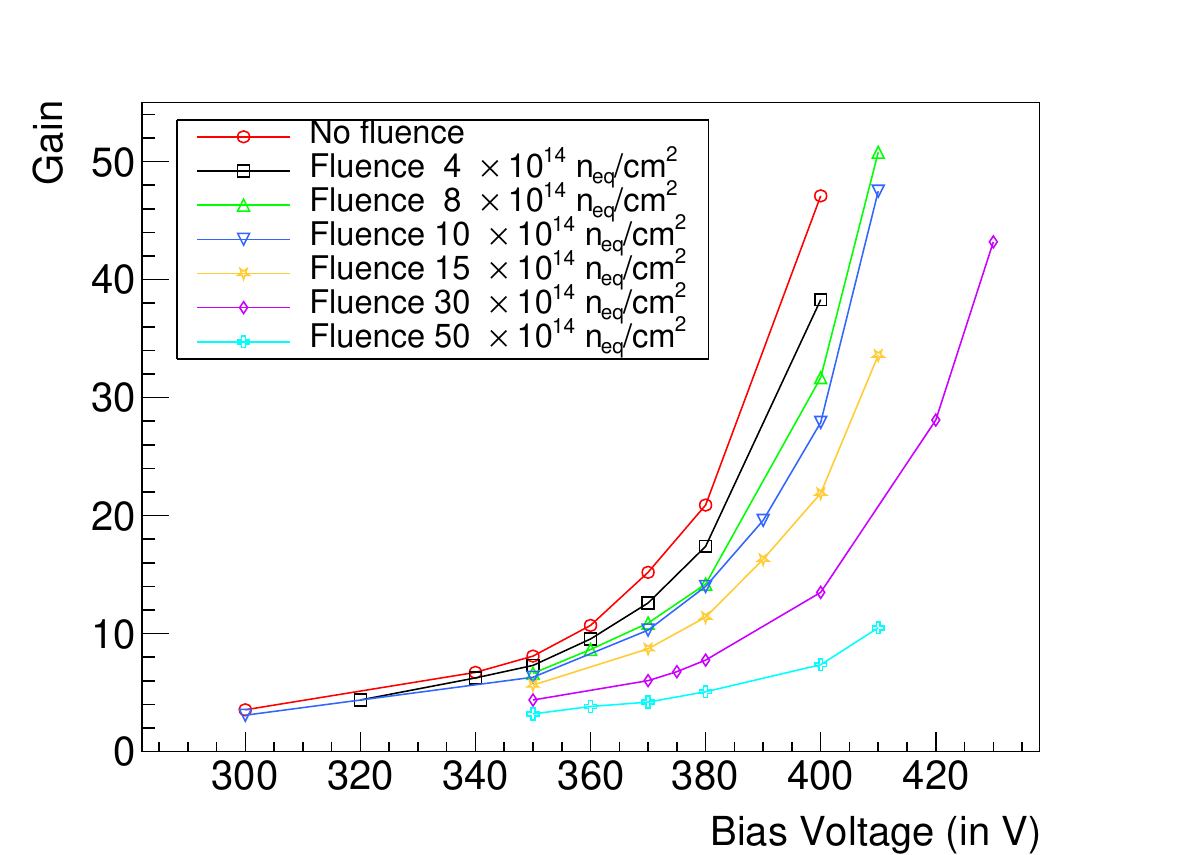} 
        \caption{Gain of the sensor as a function of sensor bias voltage at different fluences}
        \label{charge collection}
    \end{minipage}\hfill
    \begin{minipage}{0.45\textwidth}
        \centering
        \includegraphics[width=1.18\linewidth]{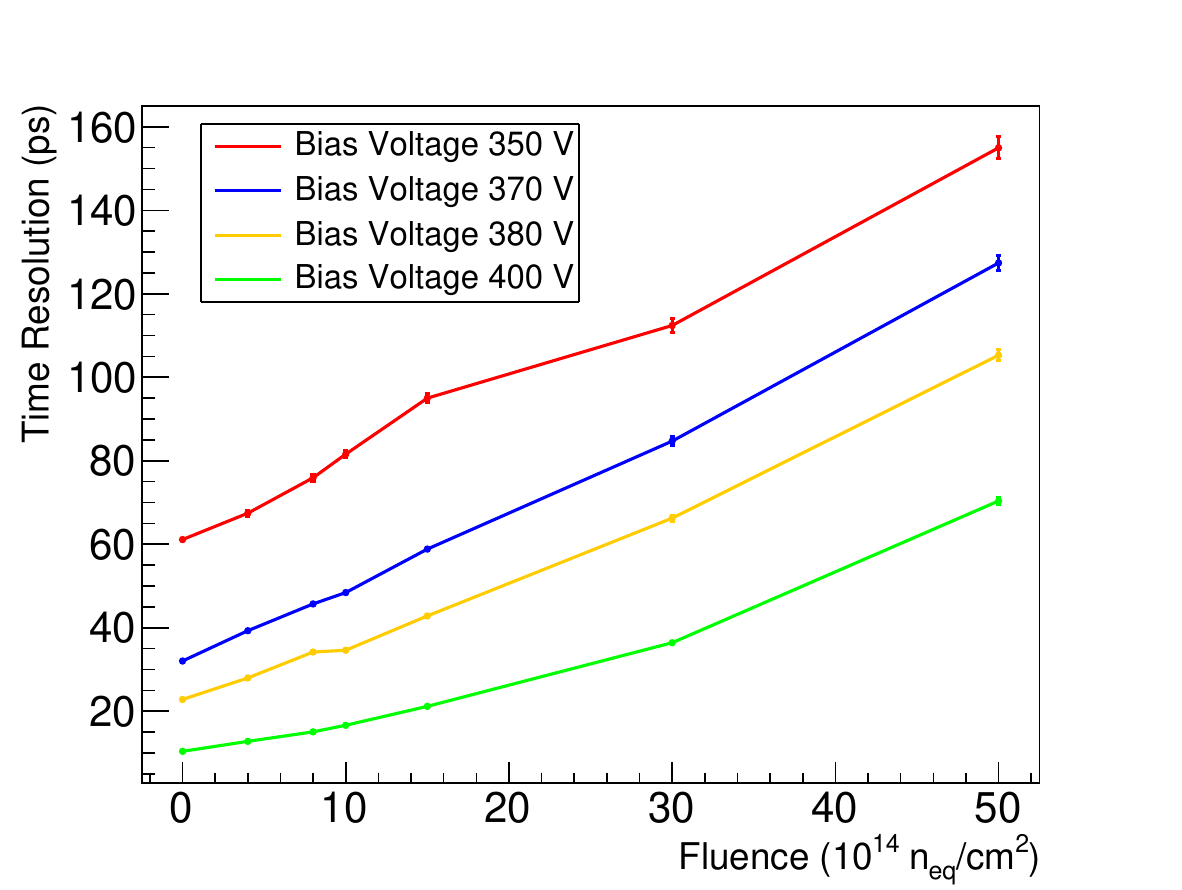} 
        \caption{Time resolution of the sensor as a function of fluence at different bias voltages}
        \label{fluence_time-res}
    \end{minipage}
\end{figure*}
\subsubsection{Temperature Related Studies}
The variation in the time resolution of a Si bulk LGAD with a thickness of 20 $\mu$m is studied with respect to increasing temperature while keeping the gain implant doping fixed at 4.01 $\times$ 10$^{16}$/cm$^3$.
With an increase in the detector thickness the time resolution increases almost linearly as can be seen in figure 3 where the gain is kept constant to see pure temperature dependence.\cite{giacomini2021,UFSD_intro}

\subsubsection{Radiation Hardness Studies}
The simulation study of Si bulk LGAD with irradiation of neutrons of different fluences is done.
The temperature of the system is kept at 293 K and the doping concentration for the gain implant layer is
kept fixed at 2.301 × 10$^{16}$/cm$^3$. The hadrons interact with the Silicon atoms of the crystal lattice, which causes surface and bulk damage. The non-combining pairs lead to point defects. This causes Non-ionising energy losses, which leads to various problems like an increase in leakage current, altering doping concentration, decreased CCE (Charge Collection Efficiency), and carrier mobility causing degraded time resolution. 
The relation between the decrease of gain with increasing fluence is shown in figure 4, where the plot is drawn between gain and bias voltage to show the effect of both on gain.
The corresponding time resolution variation for the bias voltages 350 V, 370 V, 380 V, 400 V for varying fluence is also shown in figure 5. The time resolution can be seen reaching as low as 20 ps for a high bias voltage. Also the effect of fluence can be mitigated with the increase in bias voltage. \cite{UFSD_intro,padilla2020,cenna2015} 
\subsection{Summary and Conclusions}
LGADs with thickness 20 $\mu$m - 50 $\mu$m exhibit better time resolution (20 ps) compared to the thicker ones ($>$ 50 $\mu$m). These thin LGADs can even achieve good time resolution with proper voltage conditions at high fluences. The gain of LGADs decreases as the temperature of the sensor rises, which negatively impacts the time resolution of the sensor. An increase in bias voltage can help offset the effects of radiation damage on LGADs at high fluences. The S/N ratio decreases with an increase in fluences. With significant irradiation, leakage current also increases, contributing to higher noises.

\section{Study of MPGD based MICROMEGAS using ANSYS and GARFIELD\texorpdfstring{$^{++}$}{++}}
\author{Sachin Rana, A Sarkar}


\bigskip

\begin{abstract}
Micro pattern gaseous detectors (MPGDs) represents a cutting edge technology in detection of particles which is important for applications in particle physics, security screening and  medical imaging. These detectors utilize electrodes having small scale patterns, typically on the order of tens to hundreds of micrometers, to increase the efficiency and precision of gas-based particle detection. MICRO MEsh GAseous Structure (MICROMEGAS) is a type of MPGD which offers a high spatial resolution and sensitivity. The detector consists of a two region separated by metal mesh structure called as micromesh. Control parameters of micromegas detector includes materials used,
geometry and type of gases. There are various flaws in the detector, like ion back flow, gain stability, discharge and less transparency that are needed to address. Geometry of micromegas has been constructed using simulation software
Ansys and Garfield$^{++}$ . The variation of average electric field is studied by making the different geometries and best results are obtained. Various simulations are also done to study the transparency of the detector. 
\end{abstract}
\subsection{Introduction}
By advancement in the particle accelerators, researchers are continually seeking for more efficient detectors. Gas ionization is the method used by gaseous detectors, which are especially made to measure ionizing radiation. MPGD\cite{sauli1999micropattern} is a class of advance particle detectors used in various fields in Physics. Micromegas detectors is a type of MPGD which is a high precision gaseous detector works in proportional region. Micromegas are known for their spatial resolution. In fact, among all the MPGDs, micromegas has the best spatial resolution. In this work, we are trying to replicate the geometry of micromegas with the help of various simulation software and enhancing the gain. Initial study of average electric field in the different regions is done using Ansys followed by gain analysis of the detector using Mechanical APDL\cite{thompson2017ansys} and Garfield$^{++}$. Variation of various geometrical factors like hole size, strip width and width of the avalanche region are performed using different tools and the results are presented.
\subsection{Defining the geometry of Micromegas using Ansys}
\begin{figure}[h!]
    \centering
    \begin{minipage}[b]{0.23\textwidth}
        \centering
        \includegraphics[width=3.2cm]{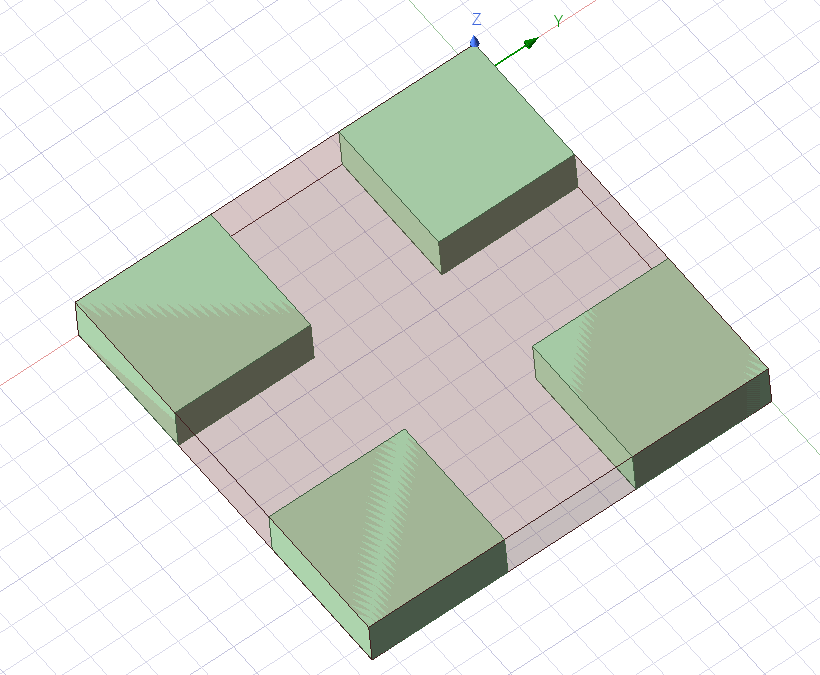} 
        \caption{Defining single unit cell}
        \label{fig:myplot9}
    \end{minipage}
    \hfill
    \begin{minipage}[b]{0.23\textwidth}
        \centering
        \includegraphics[width=3.2cm]{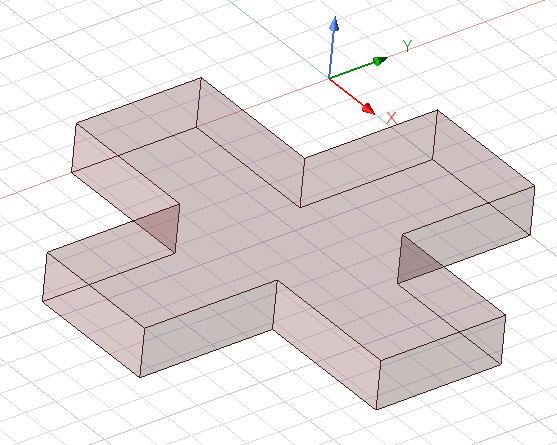} 
        \caption{Remove undesired part}
        \label{fig:myplot10}
    \end{minipage}
    \hfill
    \begin{minipage}[b]{0.23\textwidth}
        \centering
        \includegraphics[width=3.2cm]{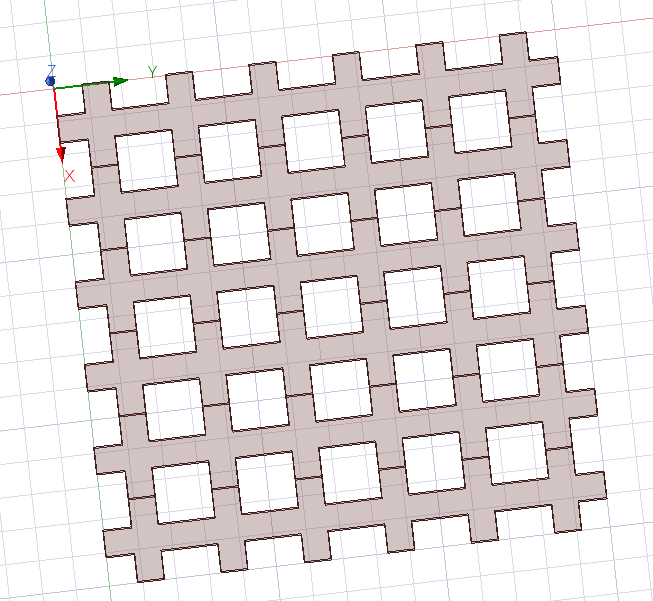} 
        \caption{Micromesh structure}
        \label{fig:myplot11}
    \end{minipage}
    \hfill
    \begin{minipage}[b]{0.23\textwidth}
        \centering
        \includegraphics[width=1.6cm, height=3.2cm]{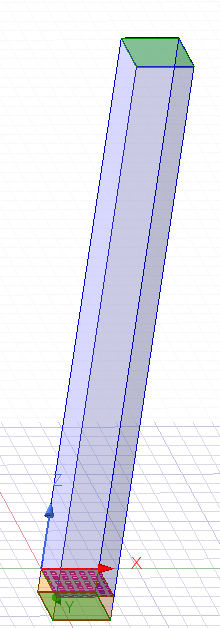} 
        \caption{Geometry of Micromegas}
        \label{fig:myplot12}
    \end{minipage}
\end{figure}

\begin{figure}[htbp]
    \centering
    \begin{minipage}[b]{0.48\textwidth}
        \centering
        \includegraphics[width=4.8cm, height=3.2cm]{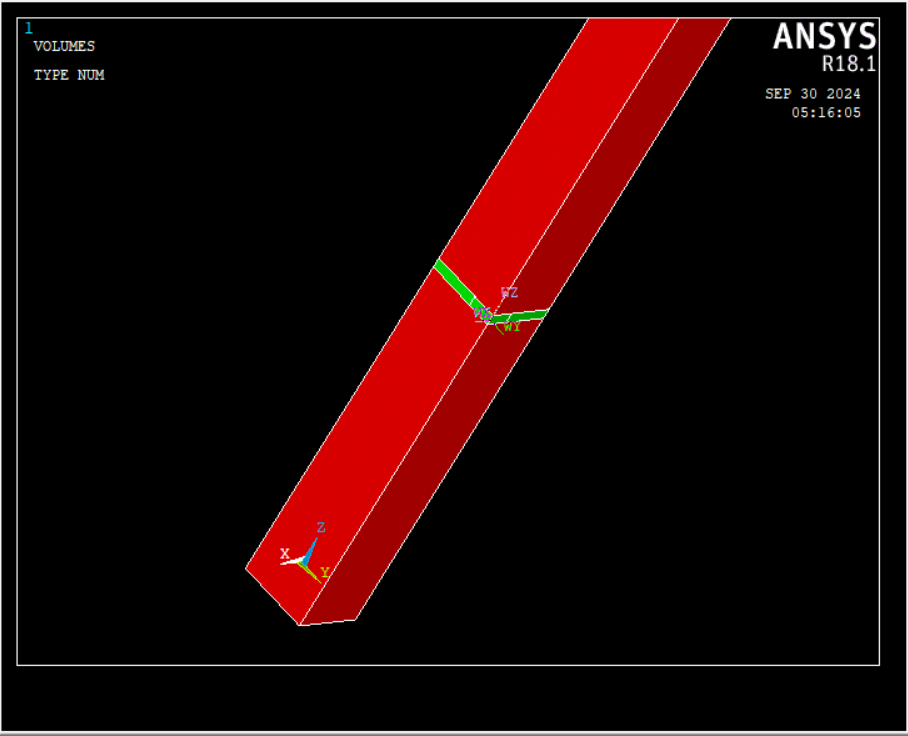} 
        \vspace{-0.2cm}
        \caption{Geometry constructed using mechanical APDL via script writing. Thin green portion is micromesh.}
        \label{fig:myplot13}
    \end{minipage}
    \hfill
    \begin{minipage}[b]{0.48\textwidth}
        \centering
        \includegraphics[width=4.8cm, height=3.2cm]{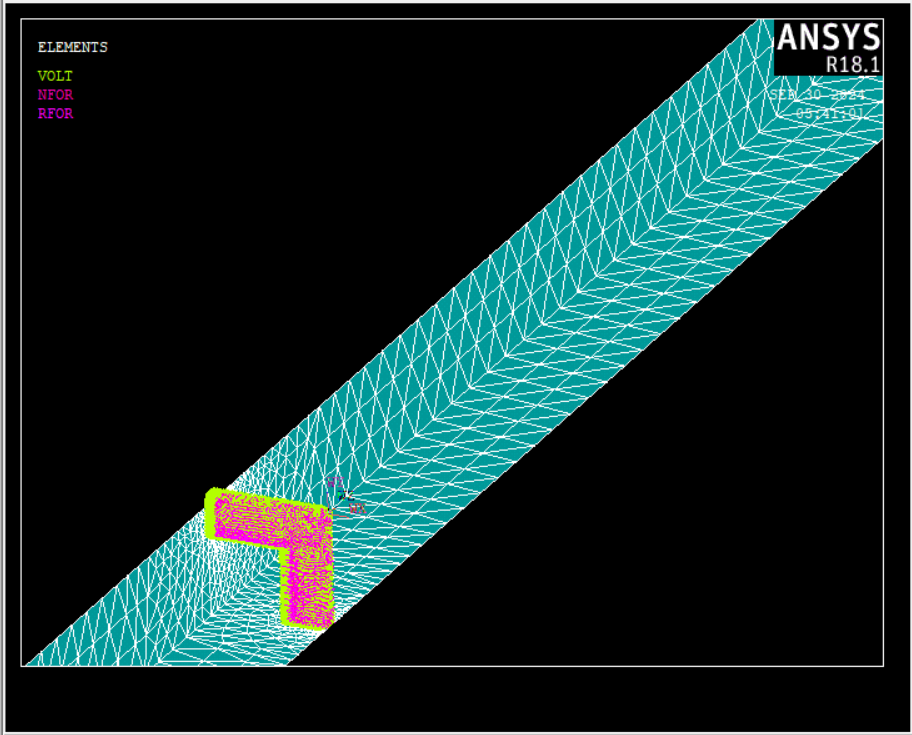} 
        \vspace{-0.2cm}
        \caption{Complete meshing of the geometry which divides the simulation domain into smaller and discrete elements.}
        \label{fig:myplot14}
    \end{minipage}
\end{figure}

To construct the geometry of micromegas, we have used the Ansys Maxwell\cite{jorgensen2020overview} software. It is a comprehensive electromagnetic field simulation software which utilize the finite element method to solve the complex electromagnetic problems with great accuracy. Step by step geometry constructed using Ansys Maxwell is shown in Figure \ref{fig:myplot9}, Figure \ref{fig:myplot10}, Figure \ref{fig:myplot11} and Figure \ref{fig:myplot12}.
Construction of geometry of micromegas starts by defining the single unit cell for mesh structure and then its repetition will make complete micromesh. This is the replica of the mesh, that can be constructed using the chemical etching process. Then it is placed at the desired location to separate drift and avalanche region. A gas mixture with 70\% Argon and 30\% CO$_{2}$ is used for the study. All study has been done without considering supporting pillars for the micromesh as we are just looking the behavior of average field for variation in different geometrical parameters. Complete geometry of micromegas is shown in Figure \ref{fig:myplot12}. Ansys Maxwell is used for the studies related to electric and magnetic field configurations. For the analysis related to gain, programming scripts were used to construct the geometry\cite{zhang2007simulation} using Ansys Parametric Design Language (APDL) which is shown in Figure \ref{fig:myplot13} and Figure \ref{fig:myplot14}.
\newline
In Figure \ref{fig:myplot13}, the small green region is the micromesh structure and the red regions below and above the micromesh are avalanche region and drift region respectively. The geometry is constructed using script writing. Figure \ref{fig:myplot14} is showing the meshing of complete geometry. It involves dividing the simulation domain into smaller, discrete elements that approximate the geometry and allow for solving physical equations over the model. After completing the construction part, Mechanical APDL generates four output files ELIST.lis, NLIST.lis, MPLIST.lis and PRSNOL.lis which contains information regarding material properties, nodes, meshing and potentials. In the next step we import these files to an another software named Garfield$^{++}$\cite{ciubotaru2018new} which is a C++ based simulation toolkit used for the optimization of gaseous ionization detectors. 
\subsection{Results and discussions}
The purpose of this work is to find the best configuration and to find the optimal parameters for the smooth working of micromegas. Firstly we observed the variation of average electric field for both drift  and avalanche region with hole size keeping strip width and the voltages at all three electrodes constant. We have found that the average electric field in drift region increase with increase in the hole diameter and average electric field is decreasing with increase in hole diameter in avalanche region. This is due to the fact that if we are increasing the hole size then there is a leakage of electric field from avalanche region to drift region.
\begin{figure}[htbp]
    \centering
    \begin{minipage}[b]{0.45\textwidth}
        \centering
        \includegraphics[width=6.4cm, height=4.2cm]{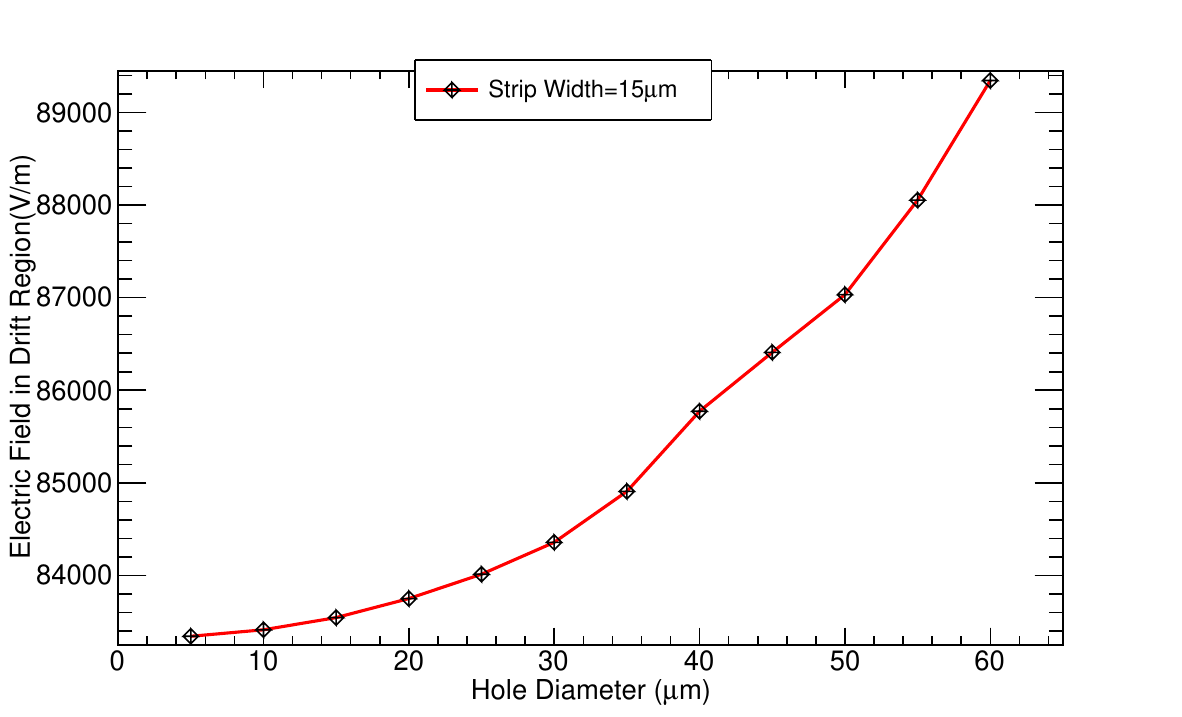} 
        \caption{Average Electric Field in Drift Region as function of Hole Size at Strip width of 15\(\mu\)m }
        \label{fig:myplot}  
    \end{minipage}
    \hfill
    \begin{minipage}[b]{0.45\textwidth}
        \centering
        \includegraphics[width=6.4cm, height=4.2cm]{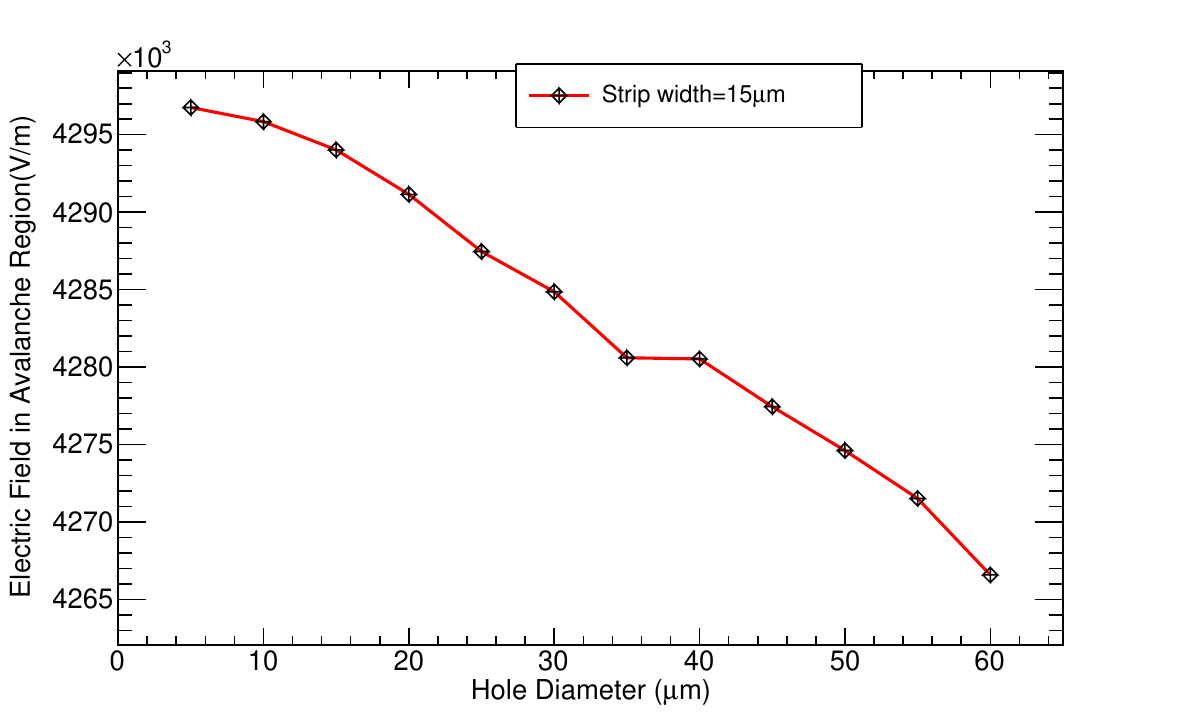} 
        \caption{Average Electric Field in Avalanche Region as function of Hole Size at Strip width of 15\(\mu\)m}
        \label{fig:myplot1}  
    \end{minipage}
\end{figure}
The behavior is shown in Figure~\ref{fig:myplot} and Figure~\ref{fig:myplot1} respectively. To confirm this behavior, same study has been done for three different strip width. The behavior of electric field for varying hole size is shown in Figure~\ref{fig:myplot2} and Figure~\ref{fig:myplot3}.
\begin{figure}[htbp]
    \centering
    \begin{minipage}[b]{0.45\textwidth}
        \centering
        \includegraphics[width=6.4cm, height=4.2cm]{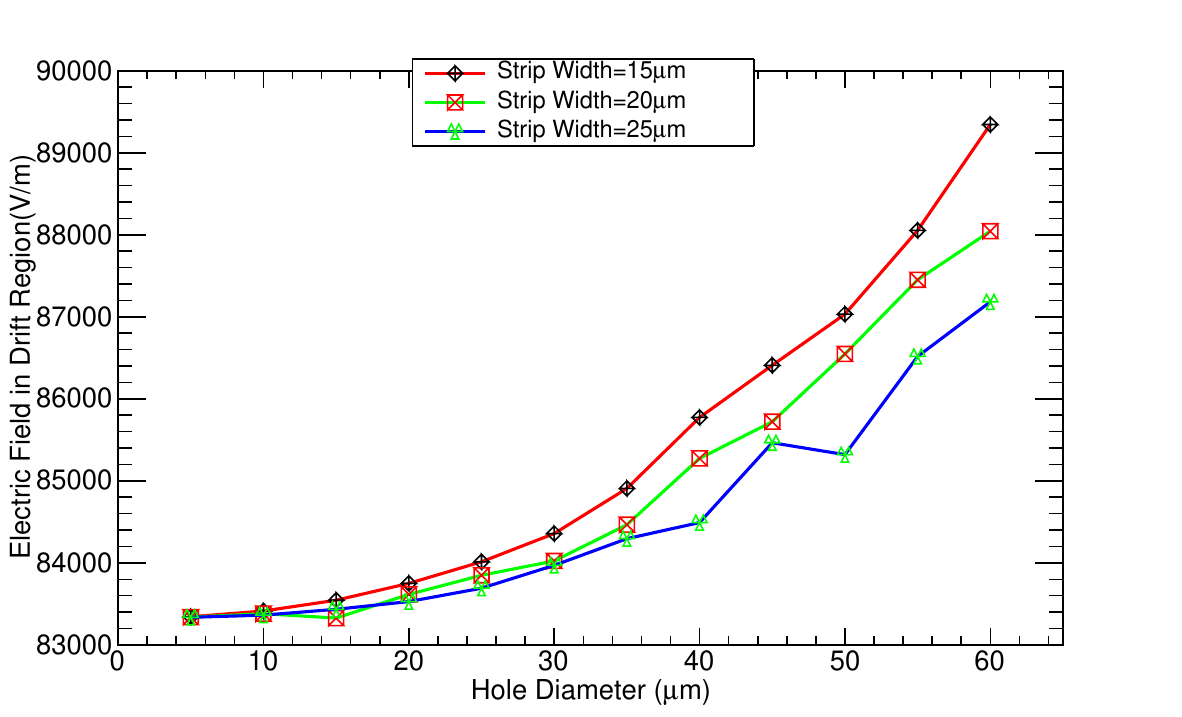} 
        \caption{Average Electric Field in the Drift Region vs Hole Size of the gaps for three different strip width. Red, green and blue lines denote the Strip width of 15\(\mu\)m, 20\(\mu\)m and 25\(\mu\)m respectively.}
        \label{fig:myplot2}
    \end{minipage}
    \hfill
    \begin{minipage}[b]{0.45\textwidth}
        \centering
        \includegraphics[width=6.4cm, height=4.2cm]{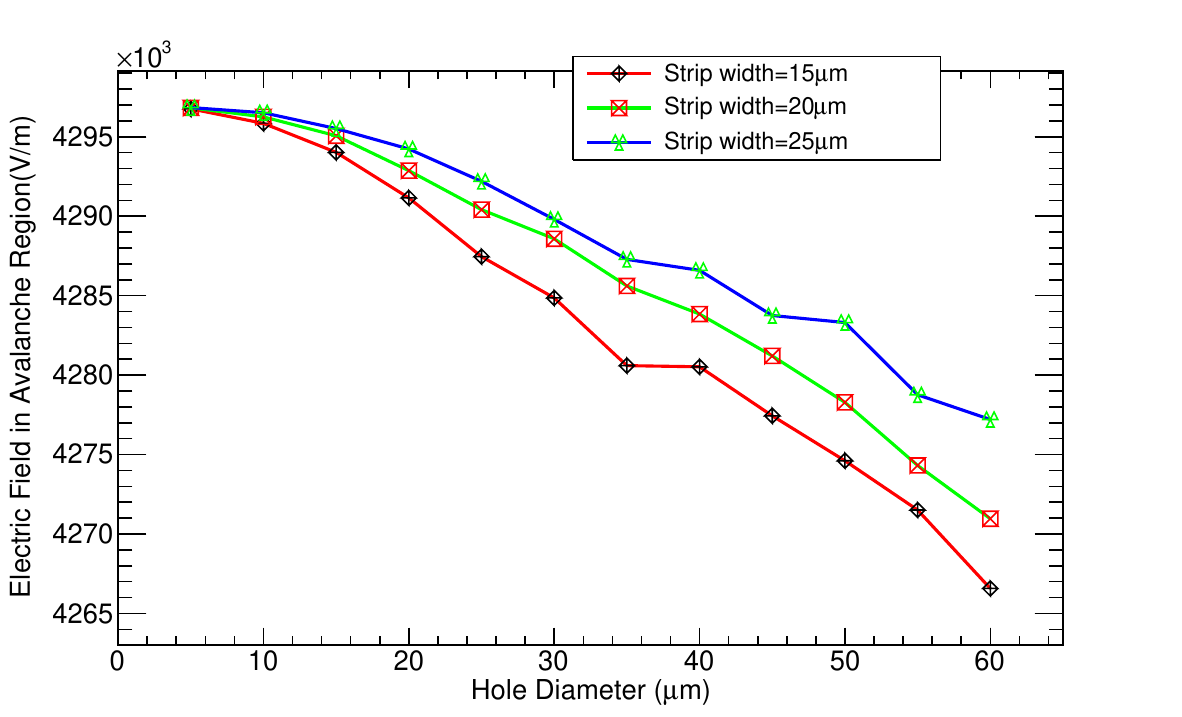} 
        \caption{Average Electric Field in Avalanche Region vs Hole Size for three different strip width. Blue, green and red lines denote the Strip width of 15\(\mu\)m, 20\(\mu\)m and 25\(\mu\)m respectively.}
        \label{fig:myplot3}
    \end{minipage}
\end{figure}
Figure~\ref{fig:myplot2} confirms that average electric field in drift region increases as we increase the hole size and Figure~\ref{fig:myplot3} gives the confirmation that average electric in avalanche region decrease as we increase the hole size. Apart from that, the behavior of average field is also studied for varying width of the avalanche region. Here width of the avalanche region refer to the available gap for the avalanche to happen i.e. the gap between the mesh and readout electrode. The result obtained is shown in Figure~\ref{fig:myplot5}.
\begin{figure}[htbp]
    \centering
    \begin{minipage}[b]{0.45\textwidth}
        \centering
        \includegraphics[width=6.4cm, height=4.2cm]{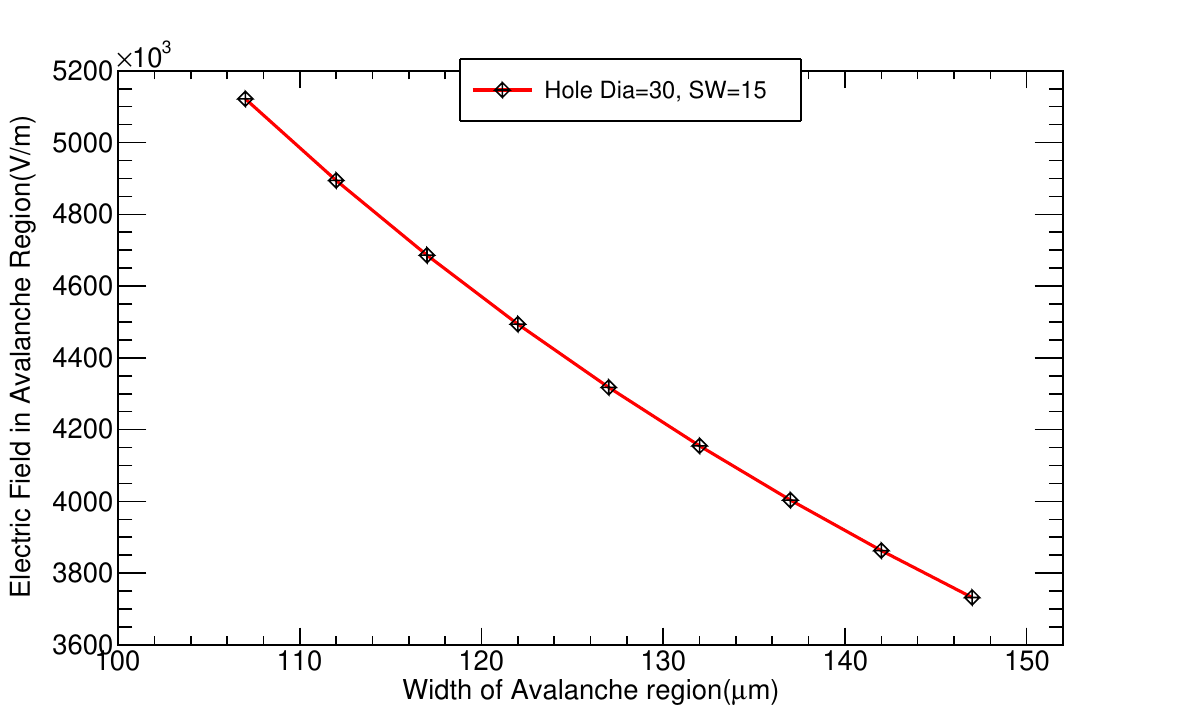} 
        \caption{Average Electric Field in Avalanche Region vs width of Avalanche region}
        \label{fig:myplot5}
    \end{minipage}
    \hfill
    \begin{minipage}[b]{0.45\textwidth}
        \centering
        \includegraphics[width=6.4cm, height=4.cm]{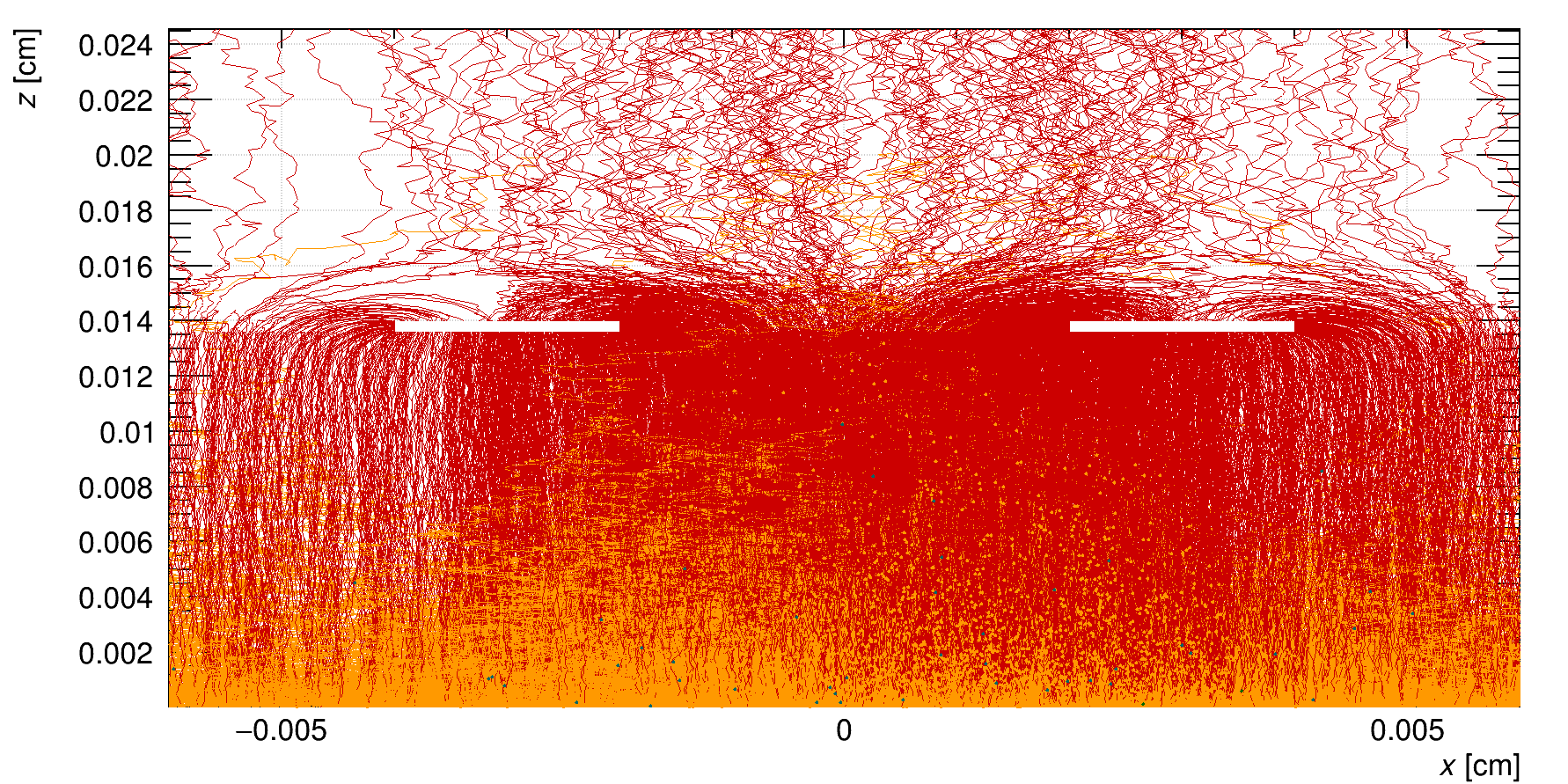} 
        \caption{Ionization process generated using Garfield$^{++}$. Red and yellow lines represent ion and electrons} 
        \label{fig:myplot6}
    \end{minipage}
\end{figure}
Smaller the width of the avalanche region small will be the gain. Here we can see that the average electric field is decreasing as we are increasing the width of the avalanche region. 
\subsubsection{Gain calculation:}
The gain for micromegas depends on the detector's  geometry as well as the environmental factors. Theoretically, it is given as:
\newline
\begin{center}
    G = e$^{ad}$ with $\alpha$=APe$^{-BP/E}$
\end{center}
where d is the width of avalanche region and $\alpha$ is first Townsend coefficient. E is given as:
\begin{center}
    E=V$_{mesh}$/d
\end{center}
A and B are the coefficient depending on the gas mixture. P being the pressure of the gas. V$_{mesh}$ is the voltage at micromesh.\\
\textbf{Gain calculation using Garfield$^{++}$}: With the help of Garfield$^{++}$, one can simulate the process of gas ionization. The avalanche process obtained using Garfield$^{++}$ is shown in Figure~\ref{fig:myplot6}. Here, in this figure the red and yellow lines are the ions and electrons produced in the avalanche process respectively. The two rectangular white strips are the part of micromesh. Red lines going above this part are the ion backflow. For the pitch size of 50$\mu$m and hole diameter of 30$\mu$m the gain value obtained using Garfield$^{++}$ is 450. 
\subsection{Summary and Conclusions:}
In this study, we examined how electric field behaves in different regions of the detector. Investigation on the variation of the hole size and strip width of the micromesh was conducted to observe their effects on the electric field. Simulations were performed using Ansys and Garfield$^{++}$ to evaluate the gain of the detector. These findings provides insights into optimizing the detector's geometry for improved performance.

\section{Study of Gas Electron Multiplier Detector Using ANSYS and GARFIELD\texorpdfstring{$^{++}$}{++}}

\author{Md Kaosor Ali Mondal, Poojan Angiras, A Sarkar}

\bigskip
\begin{abstract}
Micro-Pattern Gas Detectors (MPGDs) represent a category of gaseous ionization detectors that utilize microelectronics. They are usually filled with gases and have a remarkable tiny space between the anode and cathode electrodes, which has a considerable potential difference. Highly energetic particles can be detected efficiently by the detector. The Gas Electron Multiplier (GEM) is a type of MPGD constructed with a polyimide film sandwiched between two conductors under a high-voltage difference. Microscopic holes in the foil facilitate electron avalanche. The geometry of the GEM detectors currently in use generally exhibits suboptimal gain. Therefore, we have modified the geometry using ANSYS and conducted simulations with the assistance of GARFIELD$^{++}$ to achieve improved gain and address problems such as the back current of the detector.
\end{abstract}
\subsection{Introduction}
Gas Electron Multiplier (GEM) \cite{sauli2004progress} detectors are widely used in particle physics to track particles and analyze their behavior. Since their invention in 1997 by F. Sauli at CERN, GEM detectors have played a crucial role in detecting particles such as X-rays, alpha particles, and beta particles. These detectors belong to a class known as Micro-Pattern Gas Detectors (MPGDs) \cite{sauli1999micropattern}, which utilize a gaseous medium and electric fields to detect charged particles. Compared to older detectors such as Geiger-Müller (GM) counters, drift chambers, and multiwire proportional chambers (MWPC) \cite{gatti1979optimum}, GEM detectors offer significantly higher energy and position resolution, along with higher gain, making them particularly valuable for detailed particle studies. The core of a GEM detector consists of a thin 50~\(\mu\)m Kapton layer coated with two 5~\(\mu\)m copper layers. The GEM foil contains numerous holes with a pitch size of 140~\(\mu\)m. These detectors are filled with a mixture of Ar (70\%) and CO\textsubscript{2} (30\%) gases. When highly energetic particles pass through the gas, ionization occurs, creating electrons and ions. The electrons are then guided into the GEM holes by the applied electric field. Inside the holes, the electrons undergo multiplication through further ionization, leading to an enhanced signal. This process enables GEM detectors to detect particles with high sensitivity and accuracy.

However, GEM detectors still face some challenges. The signal strength depends on the design of the GEM foil. Another challenge is the ion backflow where ions moved backward into the detector’s drift region after an avalanche occurred inside the GEM hole, disrupting the electric field and reducing the overall performance of the detector. Ion backflow also causes unwanted noise that can interfere with the accurate detection of particles. This study aims to enhance electron gain, reduce ion backflow, and improve the durability of the detector. 
\subsection{Analysis methods}
The study begins by making several geometries using ANSYS \cite{martyanov2014ansys, thompson2017ansys}. These tools make it possible to set electrical properties and apply voltages to simulate the electric fields inside the detector. The study focuses on changing the hole shapes, and sizes to see how each factor impacts the electric field across the GEM foil. The change of gain and ion backflow was examined with the help of GARFIELD$^{++}$ \cite{Garfieldppusemanual}. This study helped evaluate how effectively the foil design can improve electron gain and minimize ion backflow.
\subsection{Results}
A new hole geometry was introduced to achieve higher gain and reduced ion backflow. Then the ratio of ion backflow to the gain for the new geometries is compared to that of the standard configuration.
\begin{figure}[htbp]
    \centering
    \begin{minipage}{\textwidth}  
        {\bf Effect of Changing Outer Hole Diameter:} \\
        The gain of the detector was calculated by varying the GEM outer hole diameter for a single electron with an energy of 0.1 eV. Figure~\ref{fig:schematic} and Figure~\ref{fig:gain-plot} illustrate the variation in gain for different outer hole diameters while keeping the inner hole diameter fixed.

        The inner hole diameters of 40~\(\mu\)m, 50~\(\mu\)m, and 60~\(\mu\)m are represented by the black, red, and green lines, respectively. The gain was maximized when the outer hole diameter was comparable to the Kapton thickness.
    \end{minipage} 

    \vspace{0.5cm} 

    \begin{minipage}{0.24\textwidth} 
        \centering
        \includegraphics[width=\linewidth]{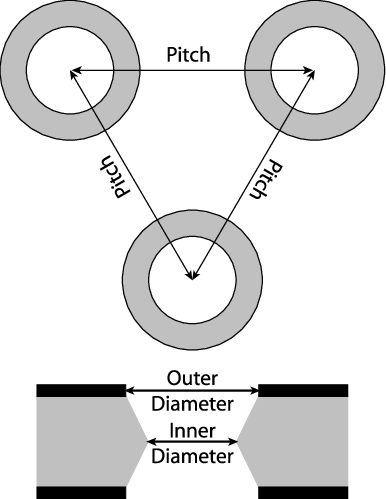}
        \caption{Schematic illustration of the geometrical parameters of GEM foils.}
        \label{fig:schematic}
    \end{minipage}%
    \hfill
    \begin{minipage}{0.50\textwidth} 
        \centering
        \includegraphics[width=\linewidth]{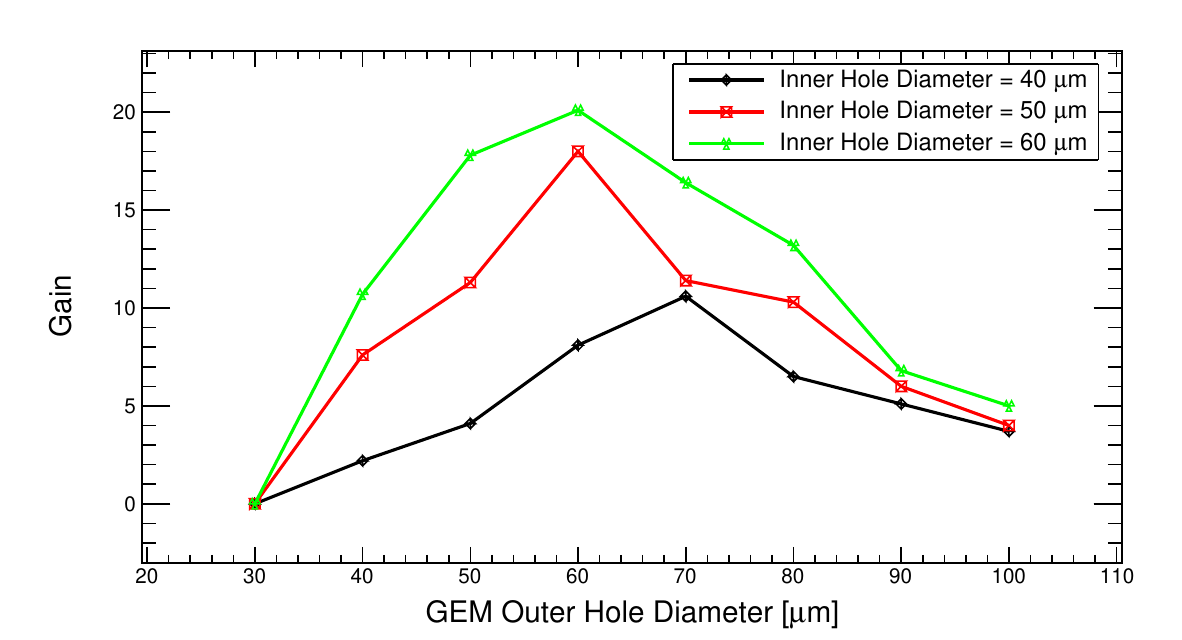} 
        \caption{Gain of GEM for various outer hole diameters with fixed inner hole diameters. Green, red, and black lines denote the inner hole diameter of 60~\(\mu\)m, 50~\(\mu\)m, and 40~\(\mu\)m, respectively.}
        \label{fig:gain-plot}
    \end{minipage}
\end{figure}

{\bf{Single conical shape hole:}}
For standard GEM holes, ions are trapped on the lower side of the kapton foil after the avalanche, which creates obstacles for further avalanche processes. The ion accumulation is decreasing the avalanche rate as well as the detector performance. A single conical-shaped hole with a 50 \(\mu\)m upper diameter and a 70 \(\mu\)m lower diameter was introduced to counter this issue; in spite of trapping ions, there is more space for avalanches. This new geometry increased the avalanche rate and efficiency. For the standard GEM and the new single conical-shaped GEM, the gain for one-line spectra is 15.6 and 2.3 respectively. 

{\bf{Variation of the lower copper thickness:}}
The uneven hole sizes on the upper and lower sides of the conical shaped holes make the foil fragile. The thickness of the lower copper layer was incrementally augmented to enhance stability. Both the copper layers collect ions, and by increasing the lower copper thickness, more ions are captured, potentially reducing ion backflow and the back current. This adjustment can alter the electric field distribution in each region of the detector, impacting both gain and ion backflow.\\
$\boldsymbol{Case I: V_{GEM}\ constant:}$ The potential difference between the copper layers is maintained at a constant value $(\Delta V_{\text{GEM}} = 300  \text{V})$. With the lower copper thickness increasing, the volume inside the hole also expands, and the electric field decreases (Eq.~\ref{eq:E_strength} ). The electric field strength \(E_{\text{strength}}\) is,

\begin{equation}
E_{\text{strength}} = \frac{\int E \, dV}{\int dV}
\label{eq:E_strength}
\end{equation}

According to the black curve in the left panel of Figure~3, the gain slightly increases as the lower copper thickness increases up to 10 \(\mu\)m, but beyond that, the gain starts to decrease. After avalanches occur, ions are collected by both copper layers. As the lower copper layer thickness increases, it absorbs more ions following the avalanche. The right panel of Figure~3 shows that as the copper thickness grows, the number of ions that backflow decreases, as represented by the black lines.\\
$\boldsymbol{Case II: The\ electric\ field\ constant\ inside\ the\ regions:}$ To achieve a higher gain, $V_{GEM}$ was adjusted to ensure a consistent electric field across all regions. As the copper layer thickness increases, so does the avalanche, leading to a rise in gain, as shown by the red line in the left panel of Figure~3. The right panel of Figure~3 indicates that the number of ions flowing back also increases, represented by the red lines. The detector's efficiency is determined by the ratio of ion backflow to gain. A reduction in this ratio improves the detector’s efficiency. 
\begin{figure}[htbp]
\centering
\includegraphics[scale=0.31]{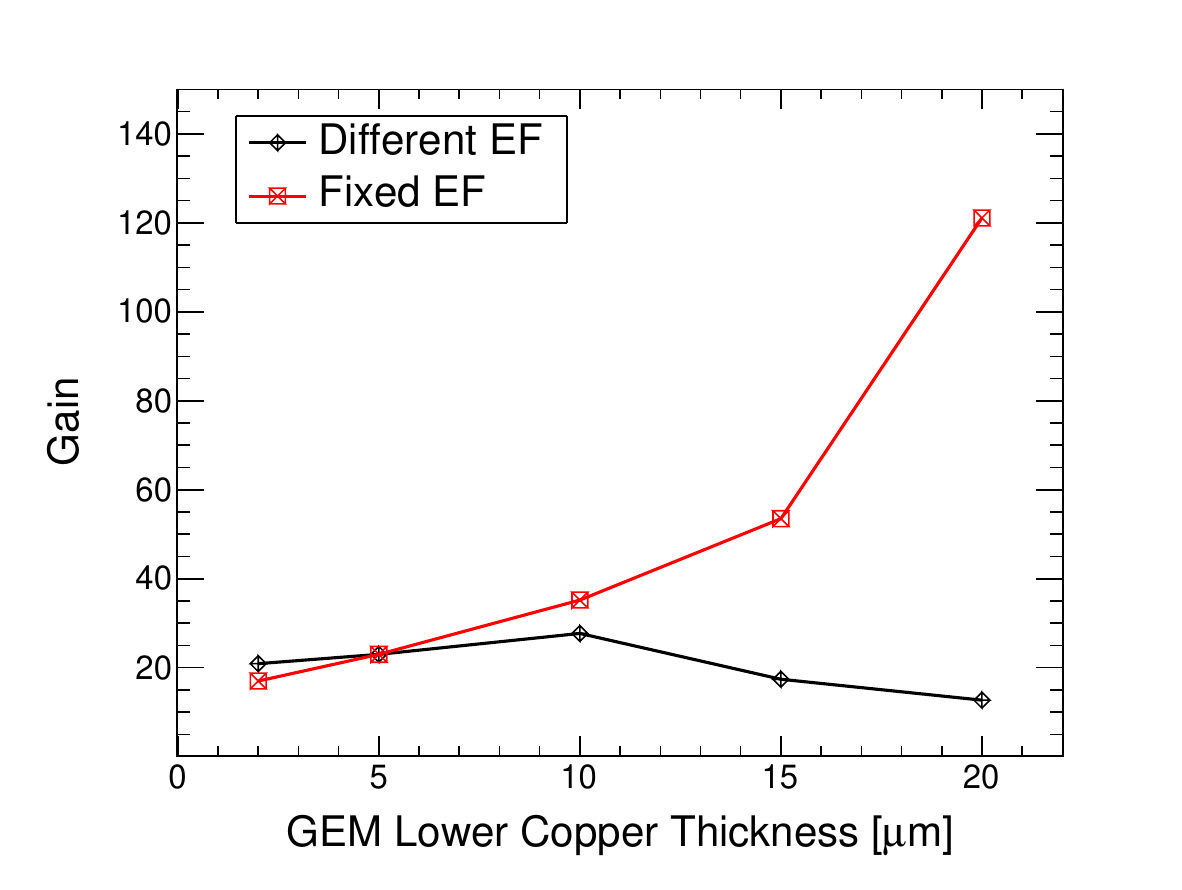}
\includegraphics[scale=0.31]{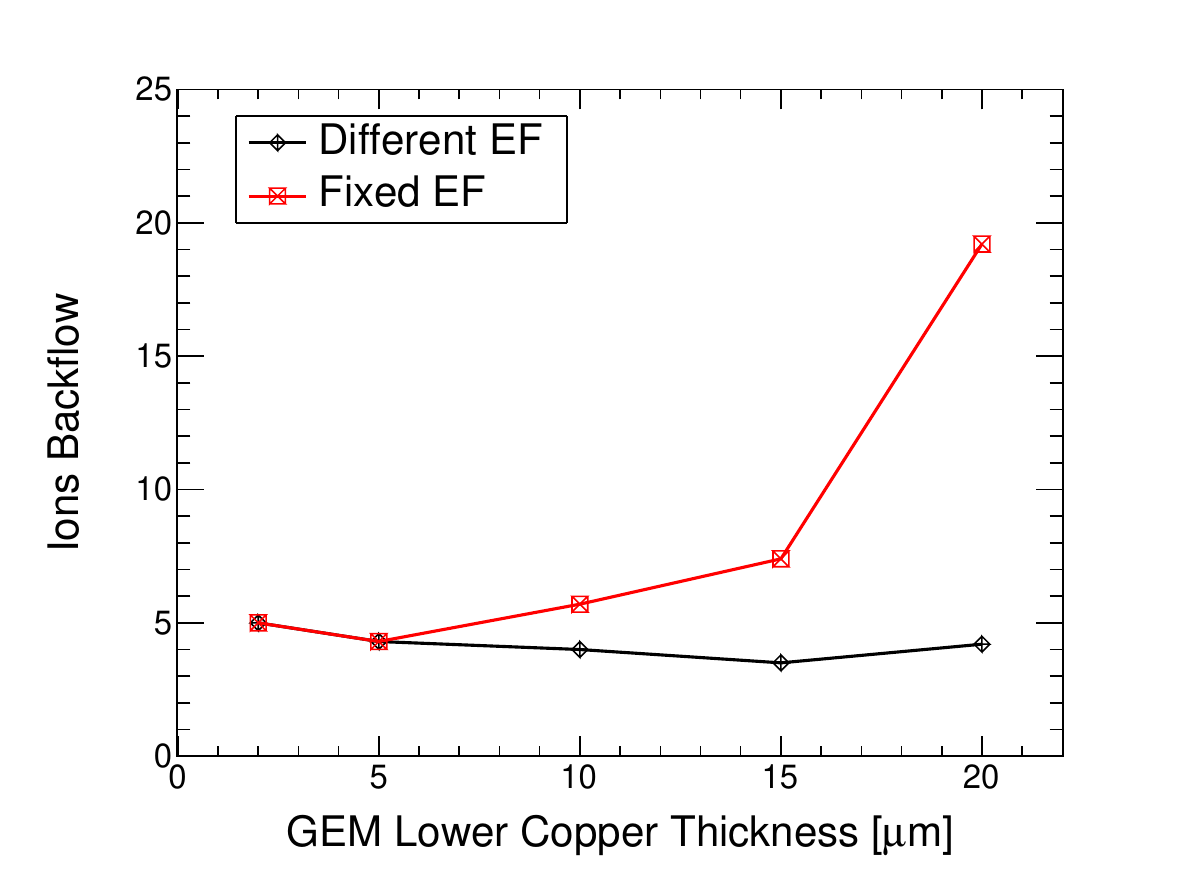}
\vspace{-0.2cm}
\caption{In both plots, the black line represents Case I: $V_{GEM}$ Constant. The red line represents Case II: Electric field Constant. The figure on the left shows how the gain changes with lower copper thickness, while the right figure illustrates the variations in ion backflow with changes in copper thickness.}
\label {Moment_products}
\end{figure}
Figure 4 illustrates how the ratio of ion backflow to gain varies with changes in the lower copper thickness. 
For Case I, the ratio decreases as the copper thickness increases up to 10 \(\mu\)m, after which it begins to rise. In Case II, the ratio consistently decreases as the copper thickness increases. For the existing GEM hole, the ion backflow-to-gain ratio was 0.2673. For the single cone design with lower copper thicknesses of 5 \(\mu\)m, 10 \(\mu\)m, 15 \(\mu\)m, and 20 \(\mu\)m, the ratios in Case I were 0.1853, 0.1437, 0.2011, and 0.3261, respectively, and for Case II, the ratios were 0.1853, 0.1609, 0.1383, and 0.1583.
\begin{figure}[htbp]
  \centering
   {{\includegraphics[width=7.2cm]{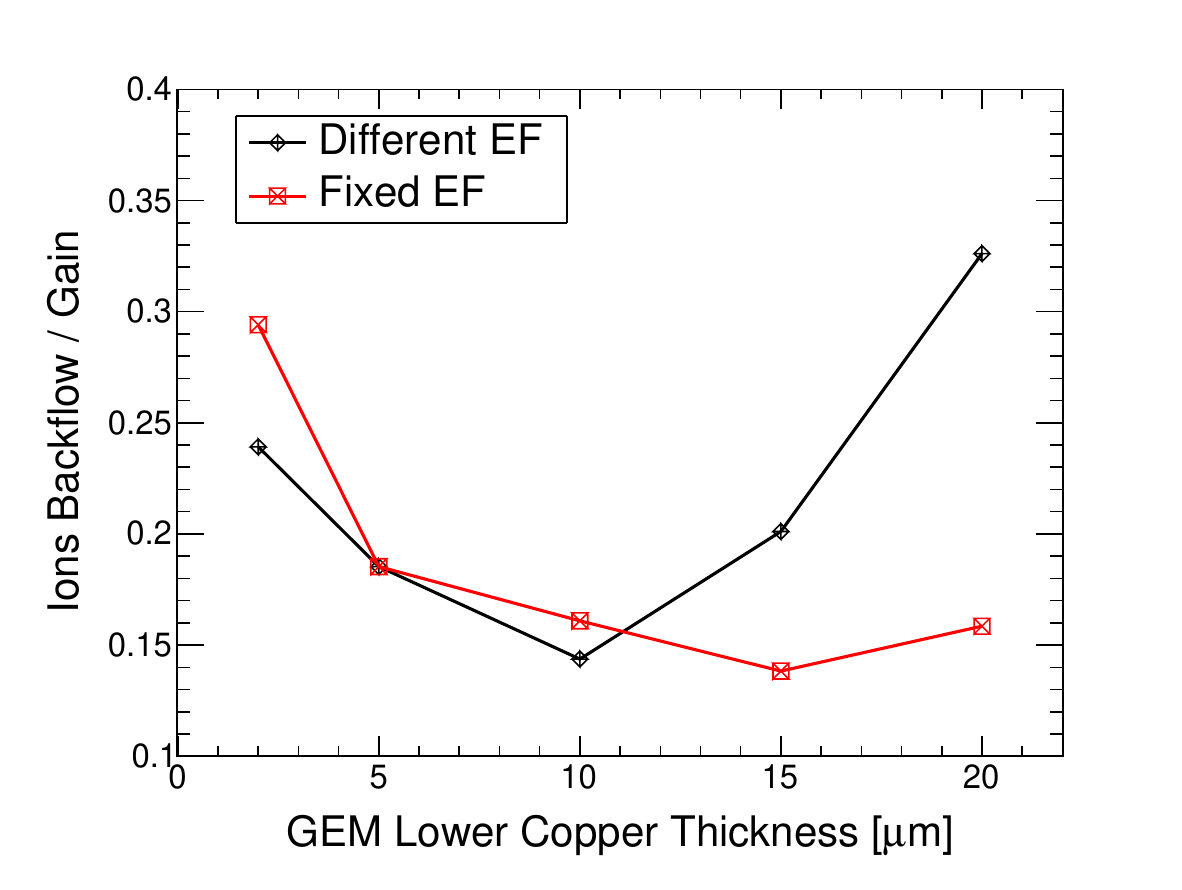} }}
   \vspace{-0.2cm}
   \caption{In Case I (black line), the potential difference ($V_{GEM}$) is constant, while in Case II (red line), it is adjusted to maintain a uniform electric field ($E$). The figure illustrates how the ion backflow-to-gain ratio changes with varying lower copper thickness} 
    \label{fig:3.0}
\end{figure}
The gain for the bi-conical and single conical shaped holes is 15.6 and 2.3, respectively, when a single electron enters the GEM hole, implying a 53\% increase in gain. A thicker copper layer was added to minimize ion backflow and improve structural durability. In Case II, the gain increased faster with copper thickness. The optimal ion backflow to gain ratio was observed to increase by 46\% for single conical holes compared to bi-conical holes for a 10 \(\mu\)m copper thickness in Case I. In Case II, the ratios increased to 40\%, 48\%  and 41\% for copper thicknesses of 10 \(\mu\)m, 15 \(\mu\)m and 20 \(\mu\)m respectively. 
\subsection{Summary}
In this study, we designed and modeled GEM detector, examining various geometries and configurations. Parameters such as pitch size, inner and outer hole diameters, and hole shape were systematically varied to assess their impact on the electric field, identifying optimal designs for efficient electron amplification. The study shows that modifying the GEM geometry and increasing copper thickness can enhance gain, reduce ion backflow, and improve the durability of GEM foils.

\section{Characterization of MALTA monolithic pixel detectors}

\author{Theertha Chembakan, Ganapati Dash, Anusree Vijay, Prafulla Kumar Behera }

\bigskip

\begin{abstract}
Future collider experiments demand pixel detectors capable of withstanding higher energy and luminosity. In response, MALTA, a novel monolithic active pixel detector, has been developed with a cutting-edge readout architecture, offering exceptional radiation tolerance, high hit rates, superior spatial resolution, and precise timing. To optimize the sensor performance prior to deployment, comprehensive electrical characterization has been conducted. This study also includes comparative DAC analyses across sensors of different thickness, providing valuable insights for further performance enhancements.
\end{abstract}


\ccode{PACS numbers:}


\subsection{Introduction}

Silicon pixel detectors play a key role in precise tracking and vertexing in high-energy physics experiments. Radiation-hard pixel detectors are essential to withstand the extreme environments at the LHC, CERN. MALTA, a promising depleted monolithic active pixel sensor (DMAPS) for HL-LHC and future colliders, offers advantages such as superior tracking, high granularity, enhanced radiation hardness, improved timing resolution, and reduced material budget compared to hybrid pixel detectors. MALTA, featuring a 512 $\times$ 512 pixel matrix, is fabricated by TowerJazz 180 nm CMOS imaging technology with three pixel variants: STD, NGAP and XDPW~\cite{ref1}. The standard (STD) process includes a low-dose n$^{-}$ layer across the p-type substrate, while NGAP and XDPW introduce a gap in the n$^{-}$ layer and a deep p-well implant, respectively. These modifications enhance radiation hardness and detection efficiency, especially at the pixel corners. The MALTA variants are produced on high-resistivity p-type epitaxial (EPI) and czochralski (Cz) substrates. This contribution presents the results of threshold study that were achieved with the next generation MALTA chip, MALTA2, on EPI wafers.

\subsubsection{MALTA2 sensor} 

MALTA2~\cite{ref1} is the second full-scale prototype designed using TowerJazz 180 nm Monolithic CMOS technology, with the dimensions of 20.2 mm $\times$ 10.1 mm and a matrix of 224 $\times$ 512 pixels, each 36.4 $\mu$m. The design focuses on enhancing radiation hardness, in-pixel charge collection efficiency, and reducing random telegraph signal (RTS) noise. MALTA2, approximately half the size of its predecessor, incorporates a higher n$^{-}$ layer doping, is radiation-hard up to 3 $\times$ 10$^{15}$ n$/cm^{2}$, and achieves timing resolution below 2 ns. Similar to MALTA, the readout is fully asynchronous.
\begin{figure}
    \centering
    \includegraphics[width=0.47\linewidth]{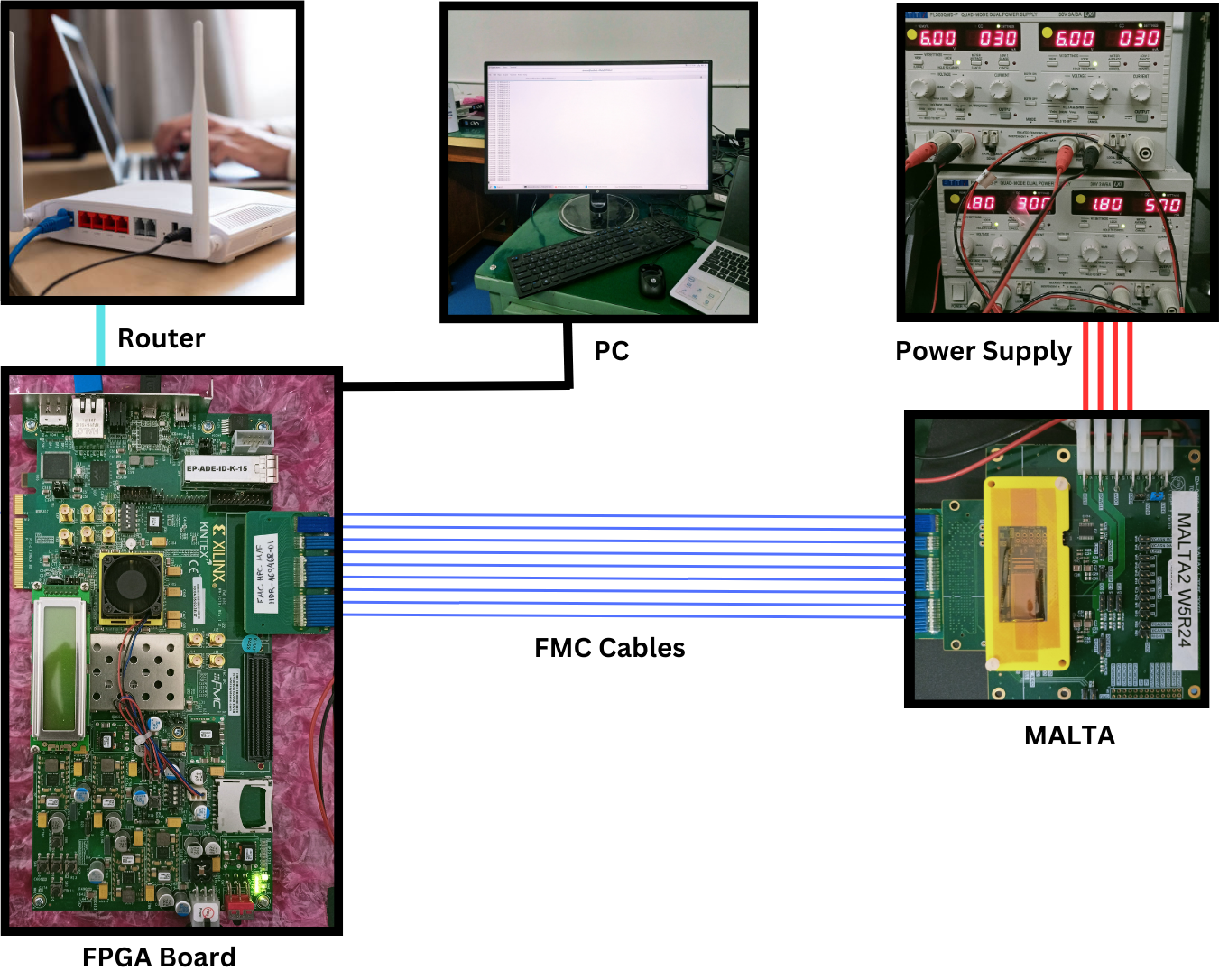}
    \caption{Malta set up at IITM }
    \label{Fig.1}
\end{figure}
\section{ MALTA Setup at IIT Madras }

To characterize the MALTA sensors, a dedicated MALTA setup has been developed in our detector development lab at IIT Madras shown in Fig.~\ref{Fig.1}. A threshold study, critical for enhancing detector performance, was conducted using this setup in an environment with temperature of 16$^{\circ}$C. The readout of the MALTA chip requires a custom carrier board, an FPGA evaluation board, and a custom firmware project. The MALTA2 chip is wire-bonded to the carrier board, which supplies various voltage levels, including LVDD, DVDD, AVDD, PWELL, SUB, and a reference voltage (DREF) for the level shifter to reset the signal~\cite{ref3}. The carrier board is connected to the FPGA evaluation board via an FMC connector. Communication is established through an Ethernet connection to a router linked to the evaluation board. Five low-voltage power supplies are used to power the MALTA chip, while a digital voltage measurement instrument is employed for calibration measurements. The MALTA software, a set of packages written in C++ and Python, is utilized for the control and readout of the MALTA CMOS prototype chips.

\begin{table}[pt]
\tbl{Description of DACs.\label{table1}} 
{\begin{tabular}{@{}cc@{}} \toprule
DAC & Description \\  
\hline
IBIAS\hphantom{00} & \hphantom{0} Main current of the Front-End, sets the power consumption. \\
IDB\hphantom{00} & \hphantom{0}Sets the threshold of the discriminator.\\
ITHR\hphantom{0} & Sets the pulse duration of the amplifier output.  \\
ICASN & Sets the baseline of the amplifier.\\ \botrule
\end{tabular} }
\end{table}

\subsection{Characterization of MALTA2 sensor}

\begin{figure}[th]
    \begin{minipage}{0.49\textwidth}
        \centering
        \includegraphics[width=\textwidth]{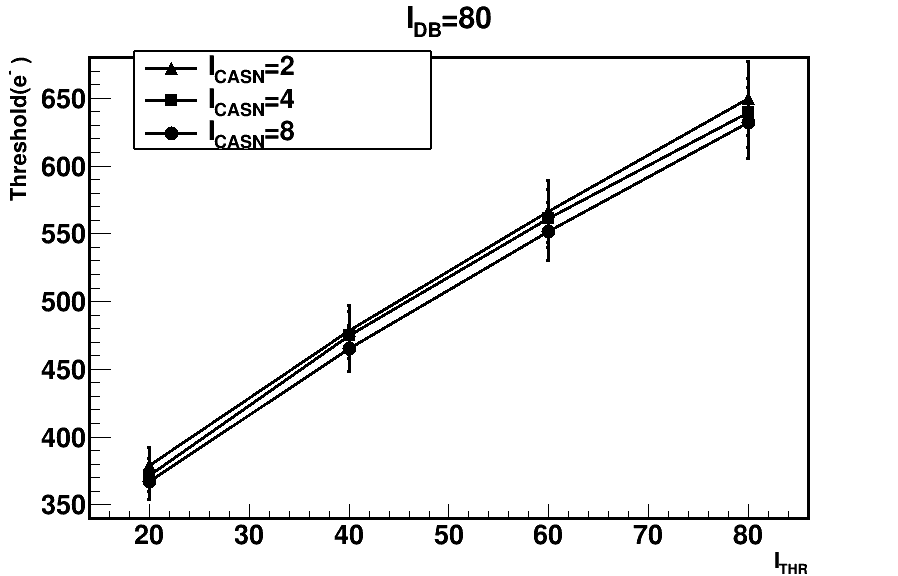}
        \caption{The distribution of ITHR for IDB = 80}
        \label{fig:2}
    \end{minipage}
    \hspace{0.005\textwidth}
    \begin{minipage}{0.49\textwidth}
        \centering
        \includegraphics[width=\textwidth]{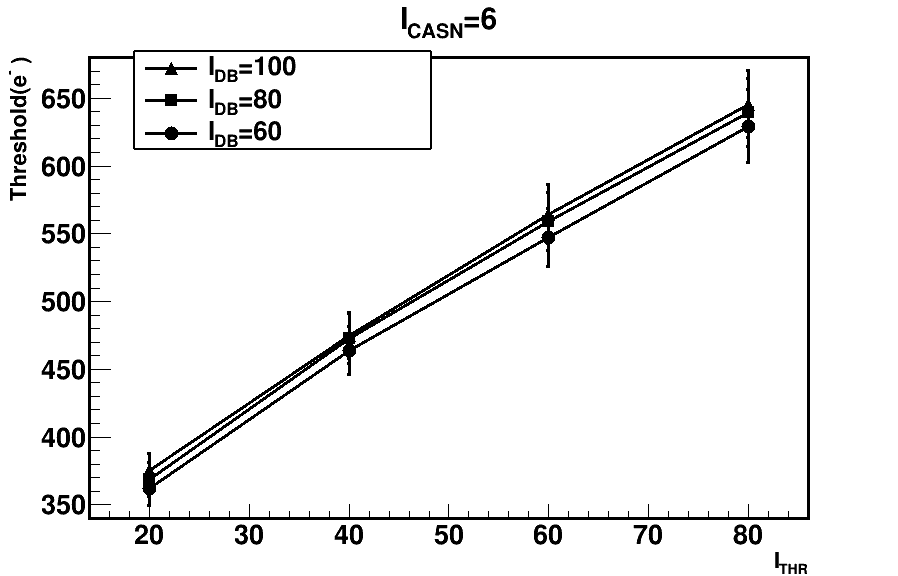}
        \caption{The distribution of ITHR for ICASN = 6}
        \label{fig:3}
        \end{minipage}
\end{figure}

\begin{figure}[th]
    \begin{minipage}{0.49\textwidth}
        \centering
        \includegraphics[width=\textwidth]{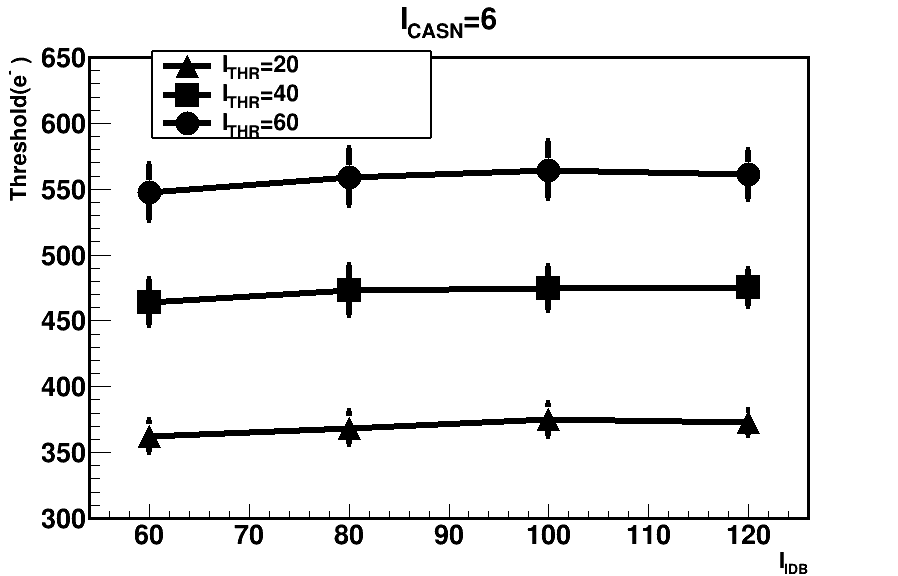}
        \caption{The distribution of IDB for ICASN = 6}
        \label{fig:4}
    \end{minipage}
    \hspace{0.005\textwidth}
    \begin{minipage}{0.49\textwidth}
        \centering
        \includegraphics[width=\textwidth]{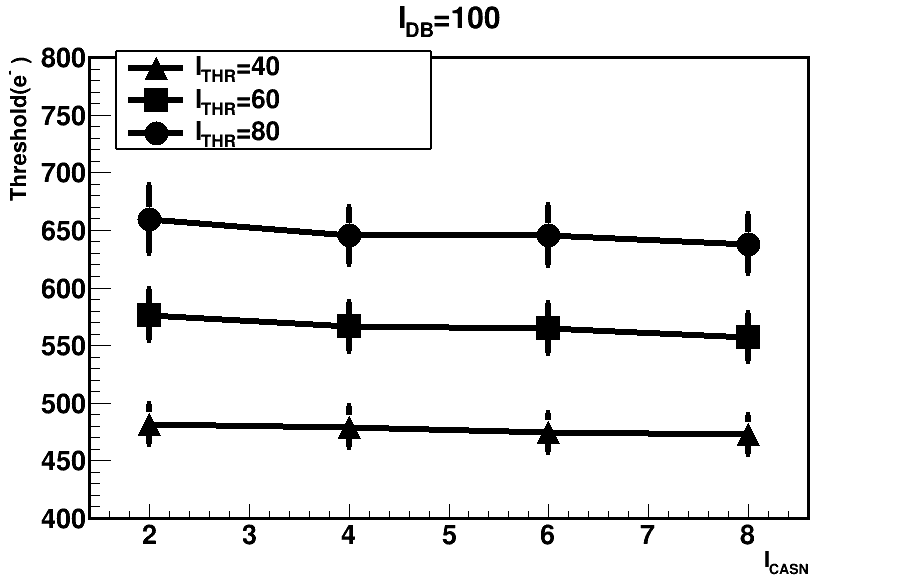}
        \caption{The distribution of ICASN for IDB = 100}
        \label{fig:5}
    \end{minipage}
\end{figure}

All properties of the fabricated detector must be thoroughly characterized and studied before its deployment in the actual detector system. To optimize sensor performance by minimizing false event triggers and enhancing detection efficiency, the threshold is a crucial parameter to evaluate. To better understand the operational range of the chip, we conducted a DAC (Digital-to-Analog Converter) study using the Malta setup at IIT Madras. The study focused on the behavior of a dedicated Malta2 chip W5R24 (NGAP, H-dope, 300 $\mu$m thickness, EPI) with a substrate voltage of -6V across various DAC settings on the sensor front end. There are eight DAC settings available for tuning the operational range of the MALTA chip: ITHR, IDB, ICASN, IBIAS, VCASN, VRESETP, VRESETD, and VCLIP~\cite{ref3} . This study specifically examined the changes in pixel threshold as three key DAC parameters—ITHR, IDB, and ICASN—were varied, while all other parameters were held constant at their standard values. The description of DACs are shown in Table~\ref{table1}. The ITHR distribution is shown in fig.~\ref{fig:2} for IDB = 80. The threshold is increasing linearly as a function of ITHR. The ITHR distribution is shown in fig.~\ref{fig:3} for ICASN = 6. The threshold increases linearly as a function of ITHR for ICASN = 6. The distribution of threshold as a function of IDB is shown in fig.~\ref{fig:4} for ICASN = 6. The threshold is constant as a function of IDB for ICASN = 6. The variation observed is within the statistical uncertainties of the data. The distribution of threshold as a function of ICASN is shown in fig.~\ref{fig:5} for IDB = 100. The threshold is constant as a function of ICASN. The variation observed is within the statistical uncertainties.
\vspace{-0.8cm}

\subsection{Conclusion}

A comprehensive threshold study was conducted across a range of DAC settings on a non-irradiated MALTA2 sample using the setup at IIT Madras. The threshold increases linearly as a function of ITHR. This is observed for different ICASN values. The threshold is constant as a function of IDB and ICASN within the statistical uncertainties. For the future, we are going to  study different thickness of sensors as well as with irradiated sensors for the detailed characterization.

\section{Characterization of proto-type silicon sensor for 
CMS Detector}

\author{Saloni Atreya, Anusree Vijay, Prafulla Kumar Behera}

\bigskip

\begin{abstract}
The silicon strip detector in the outer tracker of the CMS experiment will be upgraded for the HL-LHC run, which is expected to start in 2027. The CMS inner tracker has two sub-detectors, called the inner tracker and the outer tracker. Outer tracker is made of silicon strip sensors. Before making the specification of silicon sensors, silicon sensor proto-type were fabricated for studies. Few of the proto-type called mini sensors were sent to IIT Madras for characterization. We are involved in the characterization of mini silicon strip sensor at IIT Madras. The studies made at IIT Madras is reported.
\end{abstract}


\ccode{PACS numbers:}


\subsection{Introduction}
Silicon sensors in the CMS detector track the trajectories of charged particles and calculate their momentum. They are chosen for their high precision, radiation resistance, and low leakage current. The High Luminosity LHC (HL-LHC) upgrade \cite{CERN-LHCC-2017-009} will increase the target instantaneous luminosity by five to seven times from the present level (from \( 1 \times 10^{34} \, \text{cm}^{-2} \, \text{s}^{-1} \) to \( 5-7.5 \times 10^{34} \, \text{cm}^{-2} \, \text{s}^{-1} \)). 
The integrated luminosity will rise from 300 fb$^{-1}$ to 3000-4000 fb$^{-1}$. The number of events per bunch crossing will be increased from 35 to 140–200. We need an upgraded tracking system \cite{La_Rosa_2021} to ensure optimal physics performance in this radiation-hard environment. 
This includes enhancing radiation tolerance and detector granularity. Tracking acceptance will be expanded to \big|$\eta$\big|=4, and the tracker will contribute to the Level 1 trigger at 40 MHz as well. The trigger rate will be enhanced from 100 kHz to 750 kHz, and the material in the tracking volume will be reduced. Data will be reduced utilizing pT modules and the stub mechanism, which will filter signals below a certain threshold value\cite{Ceccarelli:2881703}. These are the requirements for the tracker in the Phase-2 upgrade to cope with high radiation and achieve the best physics performance. For these developments, proto-types of silicon sensors are fabricated by specialized semiconductor manufacturers and research institutions. In particular, we have received our sensors from Hamamatsu Photonics (Japan). We need to characterize these prototypes to ensure the accuracy of measurements.
\subsection{Results}
\begin{figure}
    \centering
    \includegraphics[width=0.5\linewidth]{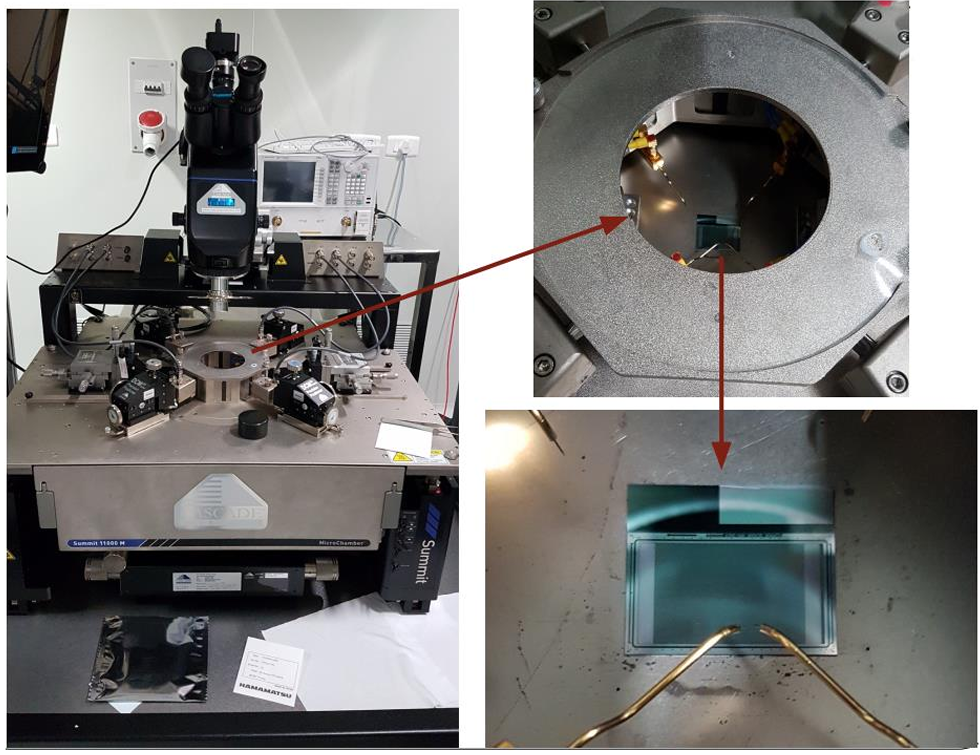}
    \caption{ Manual Probe station (Cascade Microtech Summit 11000M)}
    \label{fig: Manual Probe station (Cascade Microtech Summit 11000M)}
\end{figure}

\begin{figure}[h]
    \centering
    \begin{minipage}{0.48\textwidth}
        \centering
        \includegraphics[width=\textwidth]{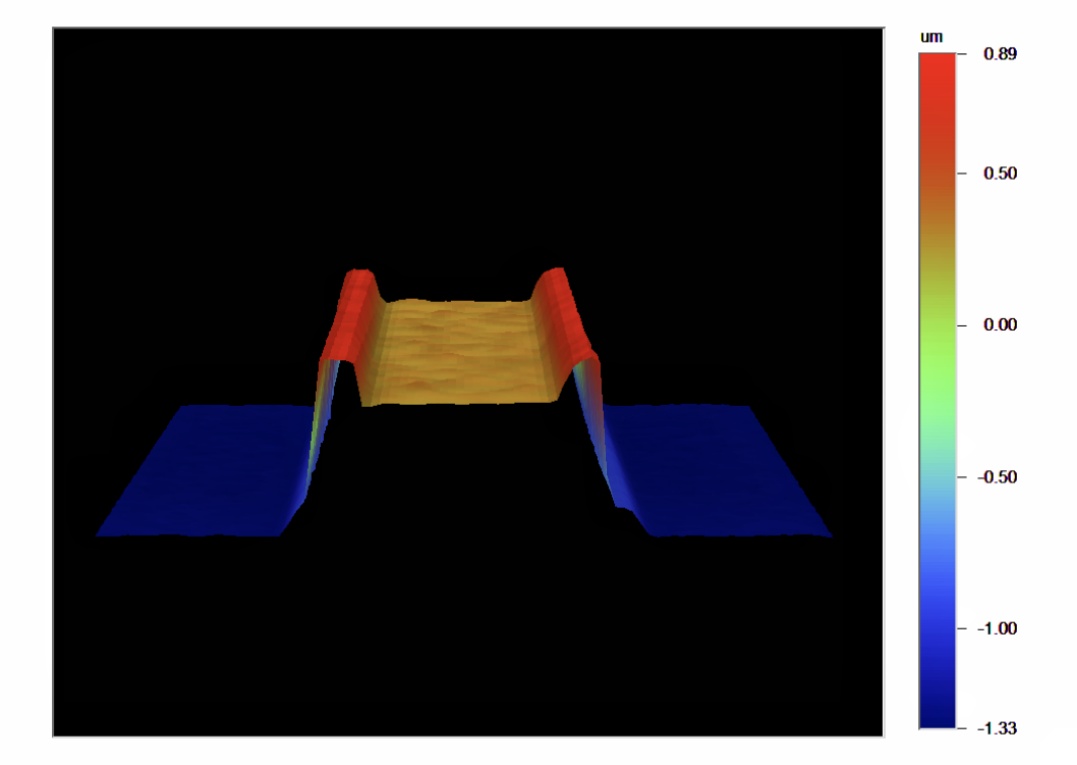}
        \caption{The single strip}
        \label{fig:The single strip}
    \end{minipage}
    \begin{minipage}{0.48\textwidth}
        \centering
        \includegraphics[width=\textwidth]{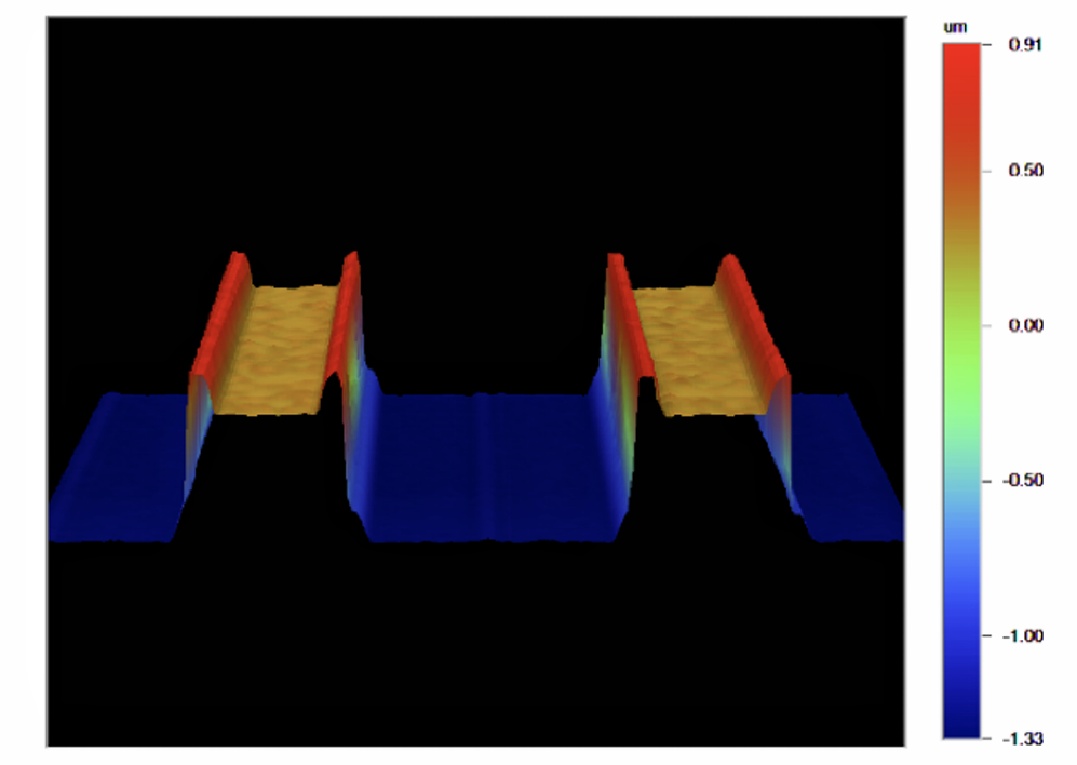}
        \caption{Gap between two strips}
        \label{fig:Gap between two strips}
    \end{minipage}
    \caption{Sensor specifications using optical surface profiler}
    \label{fig:Sensor specification using optical surface profiler}
\end{figure}

To characterize the mini-strip sensor, we have inspected for any physical damage due to transportation through the manual probe station (Cascade Microtech Summit 11000M). Upon initial inspection, there appeared to be no physical damage to the sensor. The manual probe station and micro-sensor holder is shown in Fig. \ref{fig: Manual Probe station (Cascade Microtech Summit 11000M)}.
 Next, we took certain particular measurements of the sensor for physical characterization. The width of a single strip is measured, shown in Fig. \ref{fig:The single strip}, and the measured value is 32 $\mu$m. The gap between two strips is also measured to be 58 $\mu$m, shown in Fig. \ref{fig:Gap between two strips}. The images in Fig. \ref{fig:Sensor specification using optical surface profiler} are taken using an optical surface profiler. The thickness of the wafer is measured, and that is 320 $\mu$m. The number of strips per mini sensor is 127, whereas strip height is measured to be 2 $\mu$m.\\

In the electrical characterization of these sensors, 
\begin{figure}[h]
    \centering
    \begin{minipage}{0.48\textwidth}
        \centering
        \includegraphics[width=\textwidth]{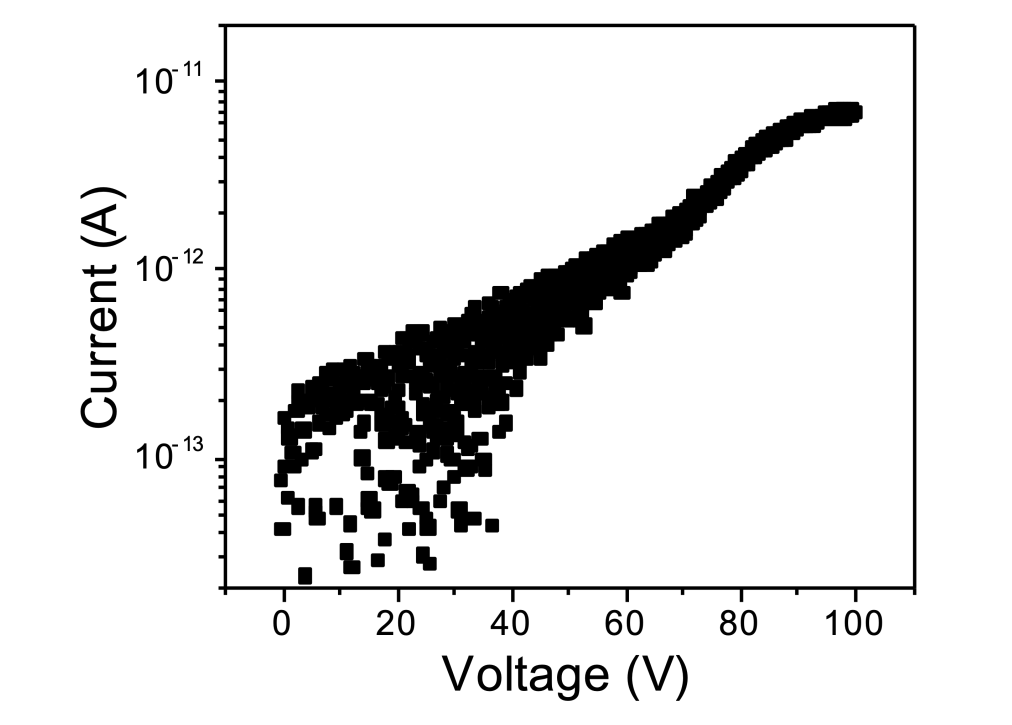}
        \caption{Measurement of current as a function of applied voltage for insulating material}
        \label{fig: Insulating material}
    \end{minipage}
   \hspace{0.1cm}
    \begin{minipage}{0.46\textwidth}
        \centering
        \includegraphics[width=\textwidth]{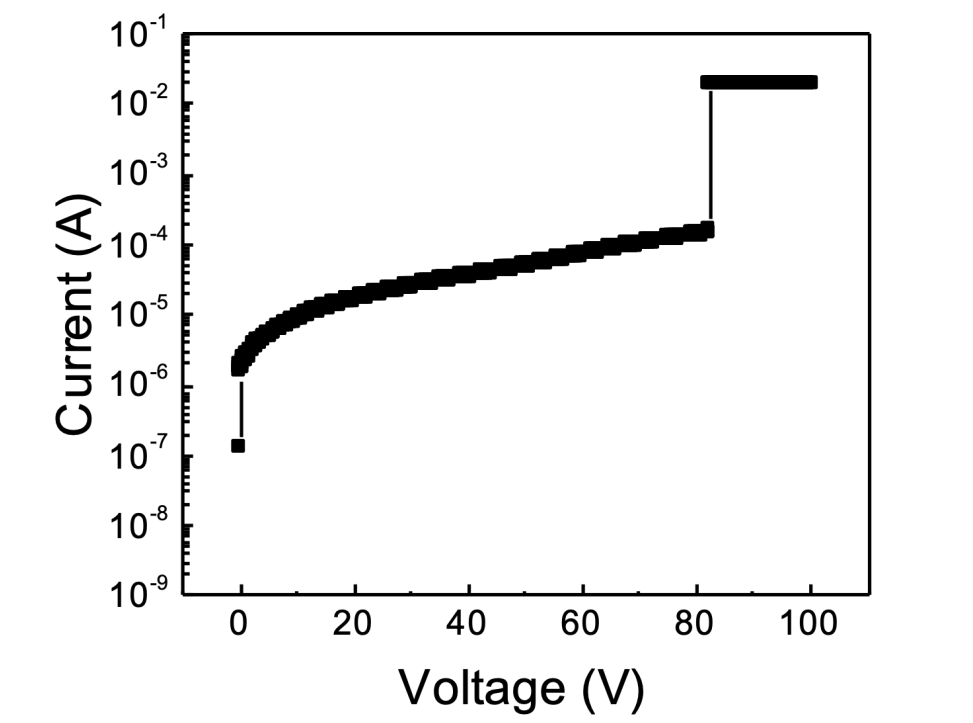}
        \caption{Measurement of current as a function of applied voltage for the strips}
        \label{fig:Strip VI ch}
    \end{minipage}
    \caption{Electrical characterization}
    \label{fig:Electrical characterization}
\end{figure}

the insulating material that separates two successive strips is measured first. The insulating property is verified by Fig. \ref{fig: Insulating material}, which shows that the current is almost negligible when the applied voltage increases to 100 V. The V-I characteristics of silicon strips are also studied. The sensors are characterized in a temperature and humidity controlled environment using the characterization setup\cite{9059697}. Figure \ref{fig:Strip VI ch} shows that the current increases linearly with applied voltage until 80 V, but after that breaks down. This is consistent with the properties of a reverse bias diode. The electrical characterization results show that the strips in the sensors are completely isolated from one another. Strips behave like a reverse bias diode until the bias voltage reaches 100 V. The measured specifications matched the original design parameters.

\section{Overview of the J-PARC muon g-2/EDM experiment}

\author{Deepak Samuel}

\bigskip



\begin{abstract}
The excellent agreement of the measured electron magnetic dipole moment with theoretical estimates has long been considered one of the significant successes of the Standard Model. In the case of muons, a recent measurement of its magnetic dipole moment at Fermilab deviated from the Standard Model prediction, thus indicating a potential discovery. Though this tension is likely to be resolved with updated theoretical calculations, it is clear that complementary experiments will be required to validate the experimental results. In this context, a new experiment called J-PARC muon g-2/edm experiment has been proposed to measure the anomalous magnetic dipole moment and electric dipole moment of the muon. Unlike the previous experiments that used electric fields to provide vertical focussing and employ the magic-gamma approach, this experiment will use ultra-cold muons, eliminating the requirement for focusing electric fields. Thus the J-PARC muon g-2/edm experiment will enable a measurement of the anomalous magnetic moment of the muon with better systematics. Summarized here is an overview of the J-PARC experiment and its current status.
\end{abstract}


\ccode{PACS numbers:}


\subsection{Introduction to muon g-2}
Classically the magnetic dipole moment of a particle with charge $q$ and mass $m$ is given as: 
\begin{equation}
    \vec{\mu} = \frac{q}{2m} \vec{S}
\end{equation}
where, $\vec{S}$ is the spin of the particle. Dirac, in his relativistic theory of the electron postulated in 1928, predicted that the magnetic dipole moment of an electron is twice the calculated value \cite{dirac1928quantum}. The anomalous magnetic moment $a$ of a particle  is defined as:
\begin{equation}
    a = \frac{g-2}{2} 
\end{equation}
where g is called the g-factor which is 2 in Dirac's theory and $a$ quantifies the deviation from 2.
Julian Schwinger, in 1948, obtained the one-loop contribution to the anomalous magnetic moment $a_e$ of the electron given by \cite{schwinger1949quantum}:
\begin{equation}
    a_e = \frac{\alpha}{2\pi}
\end{equation}
where, $\alpha$ is the fine structure constant. As on date, upto tenth order QED contributions to the electron magnetic moment have been calculated and further experiments have measured the electron g-factor with an accuracy of 0.13 ppt. The experimental values are in good agreement with the theoretical estimates.
For the muon, there are measurable contributions from QED, electroweak and strong sectors. However, the measurement of the muon's magnetic dipole moment is complicated by its short lifetime. The initial experiments that used stopped muons were statistics-limited for the same reason necessitating an intense muon beam. The Fermilab experiment uses muons produced from a proton beam, circulating in a storage ring. The anomalous magnetic moment of the muon is measured from time spectrum of the positrons produced from the decay of muons in flight.

\subsection{Measurement of $a_\mu$ in a storage ring}
The precession of the muon's spin in an electromagnetic field, with respect to its cyclotron motion is given  as \cite{bailey1979final}:

\begin{equation}
    \vec{\omega} =  -\frac{e}{m} [a_\mu \vec B - 
   {(a_\mu - \frac{1}{\gamma^2-1})}\frac{\vec \beta \times {\vec E}}{c}
    +\frac{\eta}{2}(\vec \beta \times \vec B + \frac{{\vec E}}{c})]
\end{equation}
where, $\omega$ is the spin precession frequency, $\Vec{E}$ and $\Vec{B}$ are the electric and magnetic fields experienced by muon, respectively, $\gamma$ is the Lorentz factor, and $\eta$ is the muon EDM. The measurement of $a_\mu$ relies on determining  $\omega$  in specific electric and magnetic field configurations. Any uncertainties in these fields directly impact the accuracy of the $a_\mu$ measurement. Precisely measuring the fields experienced by particles in their orbital path within the storage volume presents a significant experimental challenge. To mitigate the effects of electric field uncertainties, the Fermilab experiment employs the "magic gamma" technique. This involves adjusting the beam momentum to 3.094 GeV/c, a value that aligns $a_\mu$ closely with $\frac{1}{\gamma^2-1}$. At this momentum, $\gamma$ is about 29.3 \cite{aguillard2024detailed}.

The E-field in the Fermilab experiment is required for vertical focussing that controls the beam emittance. 
The J-PARC muon g-2 experiment eliminates the requirement for E-field by using a low-emittance muon beam produced by re-accelerated thermal muons.
Table \ref{table:compare} shows the comparison of parameters of the J-PARC experiment with the previous muon g-2 experiments at BNL and Fermilab \cite{abe2019new}.

\begin{table}[ht]
\begin{tabular}{llll}\hline
                           & BNL-E821                           & Fermilab-E989        & {J-PARC}                      \\ \hline
Muon momentum              & 3.09 GeV/$c$                       &                      & {300 MeV/$c$}                         \\ 
Lorentz $\gamma$           & 29.3                               &                      & {3}                                   \\
Lifetime $\gamma\tau_\mu$  &                                    & 64.4 $\mu$s          & {6.6 $\mu$s  }                        \\
Polarization               & 100\%                               &                      & \textcolor{purple}{50\%       }                          \\
Storage field              & B=1.45 T                           &                      & {B=3.0 T}                    \\
Radius of cyclotron motion & 7.1 m                              &                      & {333 mm }                             \\
Focusing field             & Electric quadrupole                &                      & {Weak magnetic}                  \\
Cyclotron period           & 149 ns                             &                      &{7.4 ns}                              \\
Spin precession period     & 4.37 $\mu$s                        &                      &{2.11 $\mu$s}                         \\
Number of detected e$^+$   & 5.0$\times 10^9$                  & 1.6$\times 10^{11}$   &{5.7$\times10^{11}$}                \\
Number of detected e$^-$   & {3.6$\times 10^9$}                  & {–}                    & {–}                                   \\
$a_\mu$ precision (stat.)  & 460 ppb                            & 100 ppb              & {450 ppb}                             \\
$a_\mu$ precision (sys.)   & 280 ppb                            & 100 ppb              & {$\ll$ 70 ppb}                            \\
EDM precision (stat.)      & 0.2$\times$$10^{-19}$ e $\cdot$ cm & –                    & {1.50$\times$$10^{-21}$ e $\cdot$ cm}  \\
EDM precision (sys.)       & 0.9$\times$$10^{-19}$ e $\cdot$ cm & –                    & {0.36$\times$$10^{-21}$ e $\cdot$ cm} \\ \hline
\end{tabular}

\caption{Comparison of parameters from previous experiments with the J-PARC experiment. The E989 experiment uses the same storage ring magnet from E821. Empty cells in E989 indicate the same values as in the corresponding cells for E821. }
\label{table:compare}
\end{table} 

\subsection{Production of thermal muons}
An intense beam (1 MW) of 3 GeV protons hits a graphite target to produce surface muons. These surface muons result from two-body pion decay and are monoenergetic and 100 \% polarized. These muons are transported to a Muonium production target made of silica aerogel. In this target, the muons, along with the electrons form a bound state called Muonium and eventually slow down. As these Muonium atoms diffuse out of the target, the muons are separated from the electrons through a laser-induced ionization. At this point, the muons are termed thermal muons with an average kinetic energy of about 25 meV \cite{zhang2021simulation}.

\subsection{Acceleration and injection of muons}
A LINAC further accelerates the thermal muons to a momentum of 300 MeV/c. This acceleration has to be carried out within the muon lifetime of 2.2 $\mu s$ and the LINAC is designed to contain, to the extent possible, the emittance growth of the initial thermal muon beam \cite{takeuchi2024development}.
At a beam momentum of 300 MeV/c, the average lifetime of a muon is increased to about 6.6 $\mu s$. This lifetime is a factor 10 smaller compared to the Fermilab experiment which leads to a proportional loss in sensitivity. The higher magnetic field of 3.0 T and beam intensity in the J-PARC experiment compensate for this loss to a certain extent.
The accelerated muons are injected into the storage region with a solenoidal field. The muons, aided by a kicker field, finally orbit in a region where the local magnetic field is better than 1 ppm \cite{matsushita2024demonstration}. The magnetic field is produced by a standard MRI type magnet and the field uniformity is controlled by shimming. Fixed and mapping NMR probes are used to temporally and spatially map the magnetic field in the storage region \cite{10040999}.

\subsection{Detection and tracking of positrons}
Once the muons circulate in the stable orbit, the measurement starts for a window of 33 $\mu s$. During this window, muons decay to positrons and other neutrinos. The positrons enter the detector volume placed in the centre of the storage volume. The detector comprises 40 vanes that are arranged in a cylindrical fashion. Each vane consists of 16 sensors, 8 that measure the radial position and the rest measure the axial position. The sensors are silicon strip detectors with a total of 16384 strips per sensor. In addition to the positional information, the data from the detector is read out with a time stamp of  5 ns. From the hits of the positrons in the detectors, their momentum, the position and time of decay are reconstructed \cite{nishimura2015design}. 

Under experimental conditions, approximately 6 positrons are expected every 1 ns. In the data-taking time window of 5 ns, therefore, about 30 positron tracks are expected. The identification of individual positron tracks in such pileup conditions poses a challenge. Currently, a hough-transform-based approach is used in which a candidate track is first formed from detector hits, exploiting the characteristic nearly straight trajectory of high-momentum positrons in the $\phi-z$ plane. The seed track is then extrapolated to neighbouring vanes to search for close hits. A Kalman filter is then utilized to refine the track momentum. This approach yields a track reconstruction efficiency exceeding 90\% for positrons in the 200 MeV $\leq$ E $\leq$ 275 MeV energy range, even under maximum positron rate conditions.

\subsection{Alternate strategies for track finding}
The existing hough-transform-based track-finding algorithm though efficient, requires to be computationally faster to accommodate the positron rates in high pileup conditions. The track-finding algorithm is the most time-consuming part of the analysis chain and therefore is a bottleneck in the analysis. In this regard, alternate track-finding strategies are being proposed that are faster and the same time efficient. One of the possibilities is to use Graphical Processing Units (GPU) for parallelly performing hough-transform in separate time windows. In this proposal, a single thread of a GPU consumes hits from a given time window and processes it to return the hough-space bins crossing a pre-defined threshold. Though this is expected to drastically reduce the computational time, it also with a cost. The local memory available for each thread is limited and therefore it may not be possible to push all the hits to a single thread. The shared memory can be used to overcome this limitation but the bandwidth considerations that might limit the speed of transfer between the shared memory and the thread-local memory have to be considered. Furthermore, the host-to-device and device-to-host transfer overheads can also pose a significant challenge. The work is currently in progress and this remains to be seen \cite{chetri2023gpu}. 

Another strategy that is being proposed is the use of Graph Neural Networks for track finding. In this proposal, the detector hits are first transformed to an embedded space. The properties of this space are such that the hits belonging to the same track are closer while hits from different tracks are far away as defined by a metric. Then, graphs are created using direct edges between neighbouring hits in embedded space. This is followed by an edge refinement algorithm and an edge classification algorithm that removes false edges from the graph and finally, a track labelling algorithm labels matches similar graphs and assigns a unique label. This approach has been used in other experiments and has shown promising results \cite{choma2020track}. This work is also currently in progress \cite{nandakumar2024gnn}.

\subsection{Status of muon g-2 theory}
In the Standard Model, the anomalous magnetic moment of the muon is calculated via a perturbative expansion in the fine-structure constant $\alpha$, extending upto $O(\alpha^5)$. The QED uncertainty, dominated by the unknown $O(\alpha^6$) term, is negligible. Electroweak corrections are also negligible. The most significant sources of theoretical error are the hadronic contributions, in particular, the $O(\alpha^2)$ hadronic vacuum polarization (HVP) term and the $O(\alpha^3)$ hadronic light-by-light (HLbL) scattering term. A range of theoretical approaches, including dispersive methods, lattice QCD, and effective field theories, are used to compute the hadronic computations. Of the hadronic contributions, HVP is more dominant and few concerning observations in the recent past dispute the $5\sigma$ deviation of the Fermilab measurements from the theory.  

First, in the data-driven approach, in which the $e^+e^- \rightarrow \pi^+ \pi^-$ channel dominates, the latest results from CMD3 disagree with the previous results. The updated value makes the theoretical value closer to the experimental estimates. Further in the lattice-QCD approach, the results from BMW calculations disagree with the values from the data-driven approach. This too, brings the theoretical predictions closer to the experimental values \cite{aoyama2020anomalous} \cite{theory}. 
While the final conclusion on the theoretical estimates for the anomalous magnetic moment of the muon is yet to arrive, it is very clear that a new complimentary experiment will help validate the Fermilab results.

\subsection{Timeline}
The works towards accelerator design, magnet, detector, electronics and experimental facilities are in active progress while some of them are complete. The commissioning is expected to take place in 2028 and the data-taking will commence thereafter.

\subsection{Summary}
The J-PARC muon g-2/edm experiment will complement the Fermilab experiment and validate its results of a potential discovery signal. The experiment will utilize an ultracold muon beam with low emittance and therefore will not require an E-field for focussing, unlike the Fermilab experiment. The systematic uncertainty will be therefore better than the Fermilab experiment. With most of the experimental activities in active progress or completed, the J-PARC experiment is expected to commence data-taking in 2028.

\newpage
\section*{Acknowledgments}
The organizers sincerely acknowledge the Director of IIT Mandi, the Chair of the School of Physical Sciences, the Centre for Continuing Education (CCE) Office, and the support staff of the institute for their invaluable support and assistance in organizing this conference. Their guidance and resources have been instrumental in ensuring the success of this event. Amal Sarkar gratefully acknowledges the Department of Science and Technology (DST, India) for funding this research through the EEQ/2020/000607 project and the IIT Mandi SRIC seed grant support (Ref. No. IITM/SG/ASR/118). Theertha Chembakan, Ganapati Dash, Anusree Vijay, and Prafulla Kumar Behera gratefully acknowledge the financial support from the Science and Engineering Research Board, sanction no. EEQ/2018/000679 and the Indian Institute of Technology Madras. Saloni Atreya, Anusree Vijay, and Prafulla Kumar Behera acknowledge the Science and Engineering Research Board (SERB) of the Government of India. Prabhakar Palni would like to acknowledge the support from the SERB Seminar/Symposia Scheme (File no. SSY/2024/000612) and the IIT Mandi SRIC seed grant support (Ref. No. IITM/SG/2024/01-2348). Jaideep Kalani and Prabhakar Palni would like to acknowledge the School of Physical Sciences IIT Mandi and Param Himalaya computing facility.


\end{document}